\documentclass[aps,prc,twocolumn,superscriptaddress]{revtex4-2}
\usepackage[utf8]{inputenc} % Required for inputting international characters
\usepackage[T1]{fontenc} % Output font encoding for international characters
\usepackage{lmodern}
\usepackage{graphicx}
\usepackage{xcolor}
\usepackage{amsmath}
\usepackage{dcolumn}% Align table columns on decimal point
\usepackage{bm}
\usepackage{braket}
\usepackage[hidelinks]{hyperref}% add hypertext capabilities
\newcommand{\nuc}[2]{\ensuremath{\rm{^{#1}}#2}}
\def\Be{{\nuc{9}{Be}}}
\newcommand{\uvec}[1]{{ \bm{\hat{#1}} }}
\def\a{\alpha}
\def\bmp{{\bm p}}
\def\bmq{{\bm q}}
\def\bmr{{\bm r}}
\def\Be{{\nuc{9}{Be}}}
\begin{document}

%\preprint{}

\title{The ${}^9$Be photodisintegration cross section within Cluster Effective Field Theory}


%\author{}
%\thanks{}
%\affiliation{}


\author{Y.~Capitani}
\email{ylenia.capitani@unisalento.it}
\affiliation{Universit\`a del Salento, I-73100 Lecce, Italy}
\affiliation{Istituto Nazionale di Fisica Nucleare, Sezione di Lecce, I-73100 Lecce, Italy}

\author{E.~Filandri}
\email{efilandri@ectstar.eu}
\affiliation{European Center for Theoretical Studies in Nuclear Physics and Related Areas (ECT*), I-38123 Trento, Italy}
\affiliation{INFN-TIFPA Trento Institute of Fundamental Physics and Applications, I-38123 Trento, Italy}

\author{C.~Ji}
\email{jichen@ccnu.edu.cn}
\affiliation{Institute of Particle Physics and Key Laboratory of Quark and Lepton Physics (MOE), Central China Normal University, Wuhan 430079, China}

\author{W.~Leidemann}
\author{G.~Orlandini}
\affiliation{Universit\`a di Trento, I-38123 Trento, Italy}
\affiliation{INFN-TIFPA Trento Institute of Fundamental Physics and Applications, I-38123 Trento, Italy}

\date{\today}

\begin{abstract}
A low-energy calculation of ${}^9$Be photodisintegration cross section is
presented within an $\a\a n$ cluster approach. The $\a n$ and $\a \a$ contact
interactions are derived from cluster effective field theory. The two-body
potentials defined in momentum space are regularized by a Gaussian cutoff. The
associated low-energy constants are found by comparing the calculated
low-energy $T$-matrix with its effective range expansion. A three-body
state-dependent potential is also introduced in the model.
First, the ${}^9$Be three-body binding energy is studied within the
non-symmetrized hyperspherical harmonics method. Then, the low-energy cross
section is calculated via the Lorentz integral transform method, focussing on
the dominant electric dipole transitions.
A twofold evaluation of the nuclear current matrix element is presented,
employing both the electric dipole transition operator (Siegert theorem) and
the one-body convection current operator. This approach is adopted to allow for
a discussion of the effect of the many-body currents.
\end{abstract}

%\keywords{}

\maketitle

%\section{}
%\subsection{}
%\subsubsection{}

%======================================================================== INTRO
\section{\label{sec:intro}Introduction}
%========================================================================
The tendency of nuclear systems to form states characterised by stronger
binding and increased stability results in the formation of \emph{cluster}
structures.
The four-nucleon system of $^4$He, i.e.~the $\a$-particle, is highly stable,
with a greater binding energy ($\approx 7$ MeV per nucleon) compared to other
light nuclei. Furthermore, it is relatively difficult to excite, since reaching
its first excited state requires substantial energy ($\approx 20$ MeV).
This makes the $\a$-particle well-suited to representing also a nuclear
sub-unit.

Research on nuclear clustering has a long history (see Ref.~\cite{Freer} and
references therein), which also connects with nuclear astrophysics.
The Ikeda diagram~\cite{Ikeda} is based on the idea that cluster structures in
light nuclei should emerge near the $\a$-thresholds, i.e.~the energies required
for the decay into these relevant sub-units. 
This suggested that clustering could have an impact on nuclear astrophysics,
since states that exist close to the thresholds could affect the reaction rates
of processes occurring in the burning phases of stellar
evolution~\cite{Lombardo2023}.
The triple-$\a$ process and the Hoyle state~\cite{Hoyle1954,FREER20141} are
examples of this kind.

In addition to self-conjugate nuclei, clustering can also occur in systems
composed of $\a$-particles and extra neutrons ($n$).
$\Be$ nucleus provides a three-body effective clustering system with a
Borromean structure: the whole nucleus is bound, yet its constituent subsystems
-- in this case, the $\a n$ and $\a\a$ two-body systems -- are not bound.
The threshold energy for the three-body $\a\a n$ breakup corresponds to
$B_3=1.57$ MeV, while the proton separation energy relative to the
$\a$-particle is $S_p(\nuc{4}{He}) = 19.81$ MeV. The three-body binding of
$\Be$ is therefore shallow compared to the $\a$ binding. 
This clear separation of energy scales thus provides a suitable framework for
the application of an Effective Field Theory (EFT) approach.

The three-body cluster assumption has already been used in the literature to describe $\Be$~\cite{Efroscluster,Casal,japan}, with calculations carried out by using phenomenological potentials. 
In the present study, however, we employ interactions derived from a Halo/Cluster EFT, and therefore based on a more solid theoretical foundation.

Since their introduction in nuclear physics~\cite{WEINBERG1990288}, EFTs have
had a significant impact on this field~\cite{RevModPhys.92.025004}.
Nowadays, interactions from EFTs are widely employed in \emph{ab initio}
calculations, where the Schr\"{o}dinger equation is solved for the chosen
degrees of freedom.
An EFT provides a general framework for studying the low-energy degrees of
freedom of a physical system consistently with some assumed symmetries.
Specifically, the Halo/Cluster EFT~\cite{Bert2002,Bed2003,HIGA08,Chenhalo}
represents a variant of the Pionless EFT~\cite{van_Kolck_1998,Kaplan_1998}. It
is appropriate for describing bound states and reactions of halo or cluster
nuclei.
Indeed, in certain low-energy processes involving these nuclear systems, clusters can be considered as elementary degrees of freedom alongside nucleons.

Here we are interested in the theoretical study of the $\Be$
photodisintegration reaction, $\gamma+\Be\to\a+\a+n$, in a three-body 
\emph{ab initio} approach and in the regime of low energy. 
The inverse process, including both direct and sequential reactions that
combine two $\a$ and a neutron into $\Be$, is of astrophysical relevance. In
fact, in certain astrophysical environments due to neutron star mergers or
supernova explosions, where a high number of free neutrons is present,
$\nuc{4}{He}(\a \mathit{n},\gamma)\Be$ is capable of bridging the well-known
mass gaps at $A=5$ and $A=8$, leading to a contribution to the Carbon
nucleosynthesis~\cite{OBERHUMMER2001269,SUMIYOSHI2002467,Sasaqui_2005}.
In the low-energy range, the strong Coulomb barrier makes the measurement of
this reaction quite challenging: although various experimental data sets
identify quite well the resonance peaks of interest, there is still some 
discrepancy among them~\cite{utsunew,arnold,utsuold,Burda,Goryachev}.

In our calculation of the $^9$Be photoabsorption cross section the
electromagnetic current, or transition operator, is taken as the dipole
operator. Because of the Siegert theorem~\cite{siegert}, this corresponds to
take into account the leading order effects of the one-body convection current,
derived from the free $\a$-particle Lagrangian via minimal coupling with the 
electromagnetic field, as well as the two-body currents originated from the
momentum dependent terms in the potential, 
particularly from the $P$-wave $\a n$ interaction. 
By comparing the results with those obtained using only the one-body convection
current operator, it is possible to quantify the effect due to the two-body
currents.

In order to  calculate the photoabsorption cross section the Lorentz Integral
Transform (LIT) method is used~\cite{Efros_2007}.
This approach avoids the calculation of continuum states by means of an
integral transform with a Lorentzian kernel of the relevant response function.
In this way the problem is reduced to a  bound-state-like equation, which has
the ground-state wave function of the target nucleus and the proper excitation
operator as only inputs.
The integral transform approach is used here in conjunction with the
Non-Symmetrized Hyperspherical Harmonics (NSHH)
method~\cite{Gattobigio,Deflorian}. 
The bound state problem is solved variationally, by employing a 
Hyperspherical Harmonics (HH) basis without previous symmetrization. 
The basis is defined in momentum space, as the effective $\a n$ and $\a\a$
interactions originate in this space.

Our calculation concentrates on the dominant electric dipole transitions. Because the $^9$Be ground state has a total angular momentum of $3/2$ 
and a negative parity, the following three channels are available for such $E1$
transitions: $1/2^+$, $3/2^+$ and $5/2^+$. From the inversion of the resulting
three LITs, the corresponding response functions and relative cross sections
are calculated.

The structure of the present paper is as follows. 
Sec.~\ref{sec:formalism} is devoted to the formalism of cross section, 
response function and LIT approach. 
In Sec.~\ref{sec:cEFT}  we present the general features of Cluster EFT and
define the electromagnetic current for the calculation of the electric dipole
transitions.
In Sec.~\ref{sec:GS} we explain the method used to calculate the ground state
and discuss the results obtained. 
In Sec.~\ref{sec:cross} the LIT approach to calculate the response function is
explained and the total low-energy photodisintegration cross section is
calculated, focussing first on the $1/2^+$ resonance.
Sec.~\ref{sec:cross} contains also a comparison between the results obtained
using the dipole operator and the one-body convection current.
A summary and the conclusions drawn are presented in Sec.~\ref{sec:conc}.

%============================================================= LIT method
\section{\label{sec:formalism}General Formalism}
%========================================================================
As will be explained further on, one can write the total low-energy
photodisintegration cross section as
\begin{equation}
	\sigma_i(\omega_\bmq) = 
	\frac{4\pi^2\a_\mathrm{E}}{\omega_\bmq} \left|f_i(\omega_\bmq)\right|^2 
	\frac{1}{2(2J_0+1)} 
	\sum_{\lambda=\pm 1} \sum_{M_0} R_i(\omega_\bmq)\,,\label{cross}
\end{equation}
where $R_i(\omega_\bmq)$ is the response function with excitation operator
$\hat{O}_{i,\lambda}$ and $f_i(\omega_\bmq)$ is an associate function. Furthermore, $\omega_\bmq$ represents the photon energy, 
$\lambda$ denotes the two polarizations relative to real photons,
$\a_\mathrm{E}=e^2/(4\pi)$ is the fine structure constant and  $J_0,M_0$ are
the total angular momentum and relative projection of the nuclear ground state. 
The detailed form of the response function is given by 
\begin{equation}
	R_i(\omega_\bmq)= 
	\sum_{f}\hspace{-14pt}\int\,
	 | \langle \psi_f|\hat O_{i,\lambda}(\omega_\bmq)|\psi_0 \rangle |^2
	 \delta(E_f-E_0-\omega_\bmq)\,,\label{eq:response}\\[-1ex]
\end{equation}
where $\hat O_{i,\lambda}(\omega_\bmq) $ is the spherical component of the
chosen transition operator, and $\psi_0,\psi_f$ and $E_0,E_f$ represent the
wave functions and energies of the ground and final states.
We have used the symbol $\sum\hspace{-11pt}\scalebox{0.7}{$\displaystyle{\int}$}$ to indicate the sum
over all the final states belonging to the discrete and the continuum spectrum,
as well as the sum over the angular momentum projection $M_f$.

The LIT approach that we use to determine $R_i(\omega_\bmq)$ avoids the
calculation of the continuum wave function by evaluating instead an integral
transform of $R_i(\omega_\bmq)$ with a Lorentzian kernel:
\begin{equation}
	L_i(\sigma_{R}, \sigma_{I}) = \int d \omega_\bmq \, 
	\frac{ R_i(\omega_\bmq) }%
	{ \left(\omega_\bmq-\sigma_{R}\right)^{2} + \sigma_{I}^{2}} \,.\label{2.12}
\end{equation}
In fact, one can show that for $\omega_\bmq$-independent operators this can be
represented as a norm
\begin{equation}
	L_i(\sigma_{R}, \sigma_{I}) 
	= \langle\tilde{\Psi}_i | \tilde{\Psi}_i \rangle\,,\label{2.13}
\end{equation}
where the ``LIT function'' $\tilde{\Psi}_i$ is the unique solution of the
inhomogeneous equation
\begin{equation}
	\big(\hat{H}-E_{0}-\sigma_{R}-i \sigma_{I}\big)|\tilde{\Psi}_i\rangle 
	= \hat O_{i,\lambda}  \left|\psi_{0}\right\rangle \,. \label{LITeq}
\end{equation}
The advantage with respect to solving the Schr\"{o}dinger equation 
in the continuum is that Eq.~(\ref{LITeq}) can be solved with 
bound-state type methods. 
In a final step the response $R_i(\omega_\bmq)$ is obtained 
by an inversion of the transform.

In this work we use two different forms of the transition operator 
$\hat O_{i,\lambda}$. As it will be shown in Sec.~\ref{subsec:em_current}, 
the minimal coupling  of the free Lagrangian leads in leading order to the
convection current ($\hat O_{1,\lambda}$). 
Other leading order contributions are coming from the interaction Lagrangian.
In order to include them, we use the Siegert theorem, 
which leads to the transition operator $\hat O_{2,\lambda}$. 
The exact expressions of $\hat O_{1,\lambda}$ and $\hat O_{2,\lambda}$ 
are given in Sec.~\ref{subsec:em_current}, where it will be explicitly shown
that for low photon energy they are $\omega_\bmq$-independent. The functions
$f_1(\omega_\bmq)$ and $f_2(\omega_\bmq)$ will be determined as well.

In the following we describe how to obtain the Hamiltonian $\hat H$ 
within the Cluster EFT.

%============================================================= clusterEFT
\section{\label{sec:cEFT}Cluster effective field theory framework}
%========================================================================
The effective Lagrangian for non-relativistic nucleons and $\a$-particles 
must obey the low-energy symmetries of the strong interaction, 
which are parity, charge-conjugation, time-reversal and Galilean invariance. 
At low energy, the relevant contact terms of the interaction Lagrangian density
for systems including two $\a$-particles and one neutron, 
are given by~\cite{Bed2003,Bert2002,HIGA08,Chenhalo}
\begin{align}
	\mathcal{L}_{\mathrm{int}}
	&= \tilde{\lambda}^{S}_{0,\a\a} (\Psi\Psi)^\dag(\Psi\Psi) 
	\nonumber\\
	&+ \tilde{\lambda}^{S}_{1,\a\a} \left[%
	(\Psi\overleftrightarrow{\nabla}^2\Psi)^\dag(\Psi\Psi) + \mathrm{H.c.}
	\right]
	\nonumber\\
	&+ \tilde{\lambda}^{S}_{0,\a n} (\Psi n)^\dag (\Psi n) 
	+ \tilde{\lambda}^{S}_{1,\a n} \left[%
	(\Psi\overleftrightarrow{\nabla}^2  n)^\dag (\Psi n) + \mathrm{H.c.}
	\right]
	\nonumber\\
	&+ \tilde{\lambda}^{P}_{0,\a n} (\Psi\overleftrightarrow{\nabla}n)^\dag
	\cdot (\Psi\overleftrightarrow{\nabla}n)
	\nonumber\\
	&+ \tilde{\lambda}^{P}_{1,\a n} \left[%
	(\Psi\overleftrightarrow{\nabla}^2 \overleftrightarrow{\nabla} n)^\dag \cdot (\Psi\overleftrightarrow{\nabla} n) + \mathrm{H.c.}
	\right] + \dots\,,\label{Lint}
\end{align}
where $\overleftrightarrow{\nabla}$ indicates the left and right derivative 
acting as
\begin{equation}
	\Psi_{a} \overleftrightarrow{\nabla} \Psi_{b} = 
	\Psi_{a}\frac{m_{b} \partial_{a}-m_a\partial_{b}}{m_{a}+m_{b}} \Psi_{b}\,,
\end{equation}
while $\Psi$ and $n$ represent the non-relativistic $\a$-particle and 
neutron fields, respectively.
$\tilde{\lambda}^{S}_{i,\a\a}$, $\tilde{\lambda}^{S}_{i,\a n}$ and
$\tilde{\lambda}^{P}_{i,\a n}$ with $i=0,1$ in Eq.~(\ref{Lint}), 
are the low-energy constants (LECs) for inter-particle interactions 
in the $\a\a$ $S_{0}$, $\a n$ $S_{1/2}$ and $\a n$ $P_{3/2}$ channels.

From the Lagrangian~(\ref{Lint}) the $\a\a$ and $\a n$ potentials 
can be obtained as a function of the relative momentum of the two particles. 
The two-body potentials have the following form
\begin{equation}
	V(\bmp,\bmp') 
	\equiv \langle\bmp|V|\bmp'\rangle 
	= \sum_{\ell=0}^{\infty} (2\ell+1) V_{\ell}(p,p') P_{\ell}(\uvec{p} 
	\cdot \uvec{p}')\,,
\end{equation}
where $\bmp$, $\bmp'$ are relative momenta and $P_{\ell}$ is the $\ell$-th
Legendre polynomial. 
By regularizing the ultraviolet divergences with a Gaussian cutoff
\begin{equation}
	g(p)=e^{-(p/\Lambda)^{2m}}\,,
\end{equation}
with $m$ positive integer, $V_{\ell}(p,p')$ is given by
\begin{equation}
	V_\ell(p,p')
	= p^{\ell} p'^{\ell} g(p) g(p') \sum_{i,j=0}^{1} 
	p^{2i} \lambda_{ij} p'^{2j}\,,\,
	\lambda=\begin{pmatrix}
		\lambda_0&\lambda_1\\
		\lambda_1&0
	\end{pmatrix},\label{eq:partial}
\end{equation}
in which the dependence of $\lambda_{0},\lambda_{1}$ on $\ell$ is understood.
The choice of the cutoff regularization is due to its ability to reproduce
known features, such as parameters in the effective-range expansion (ERE) of
the $T$-matrix or the correct scaling of the renormalized scattering
amplitude~\cite{Phillip,Phillip2}.

%------------------------------------------------------------------------
\subsection{\label{subsec:LEC}Determination of the low-energy constants}
%------------------------------------------------------------------------
For each two-body subsystem, the various LECs are determined by expanding the
corresponding Lippmann-Schwinger equation in partial waves.

In the $\a n$ case, the Lippmann-Schwinger equation takes the form
\begin{equation}
	T(\bmp,\bmp';E)
	= V(\bmp,\bmp') + \int \frac{ d^3\bmq }{ (2\pi)^3 } V(\bmp,\bmq) 
	\frac{ T(\bmq,\bmp';E) }{ E-\frac{q^2}{2\mu_{\a n} } + i\varepsilon }
	\,,\label{eq:LSan}
\end{equation}
where $\mu_{\alpha n}$ is the $\a n$ reduced mass and on-shell, i.e.~$p'=p=k$, 
the energy is $E=k^2/(2\mu_{\a n})$.
In order to solve~(\ref{eq:LSan}), we use Eq.~(\ref{eq:partial}) and, 
similarly to what is done for the potential, we also expand the $T$-matrix as
\begin{equation}
	T(\bmp,\bmp')
	=\sum_{\ell=0}^{\infty} (2\ell+1) T_{\ell}(p,p') P_{\ell}(\uvec{p}\cdot\uvec{p}')\,,
\end{equation}
where each partial wave component $T_{\ell}(p,p')$ is parametrized as follows
\begin{equation}
	T_\ell(p,p')
	= p^{\ell} p'^{\ell} g(p) g(p') \sum_{i,j=0}^1 p^{2i} 
	\tau^{(\ell)}_{ij}(E) p'^{2j}\,.\label{eq:partial2}
\end{equation}
This allows to obtain a first expression of the on-shell $T$-matrix, 
which can be expanded in powers of $k/\Lambda$ 
and evaluated for the relevant partial waves.
Then, the LECs $\lambda_0,\lambda_1$ are determined from the term-by-term 
comparison between the calculated on-shell $T$-matrix and the following ERE 
up to order $k^2$
\begin{equation}
	\frac{ k^{2\ell} }{ T^{\mathrm{on}}_{\ell,\a n}(E) } 
	= \frac{\mu_{\a n}}{2 \pi} \left(%
	\frac{1}{a_{\ell}} - \frac{1}{2} r_{\ell} k^{2} + O(k^{4}) + i k^{2 \ell+1}
	\right)\,. \label{eq:ERE-an}
\end{equation}
This is valid for both $\a n$ $S_{1/2}$ and $\a n$ $P_{3/2}$ channels, 
taking $\ell=0,1$, respectively. 
Having required the EFT to reproduce the low–energy $T$-matrix ERE, 
the LECs are fixed on the experimental values of scattering length 
(or scattering volume) and effective range, for each given cutoff $\Lambda$. 
The experimental scattering parameters used in the present work 
are listed in Tab.~\ref{tab:ar_exp}.
\begin{table}[tbp!]
	\caption{Experimental values of the low-energy scattering parameters relative to $\a n$ and $\a\a$ systems, taken from Ref.~\cite{ARNDT1973429} and Refs.~\cite{AFZAL,HIGA08}, respectively.}\label{tab:ar_exp}
	\begin{ruledtabular}
		\begin{tabular}{cccc}
			&$\ell_j$	&$a^{\text{exp}}_\ell [\textrm{fm}^{2\ell+1}]$	&$r^{\text{exp}}_\ell [\textrm{fm}^{-2\ell+1}]$\\
			\colrule
			$\a n$	&$S_{1/2}$ 	&2.4641	&1.385	\\
			$\a n$	&$P_{3/2}$	&-62.951&-0.8819	\\
			$\a\a$	&$S_0$		&-1920	&1.099	\\
		\end{tabular}
	\end{ruledtabular}
\end{table}

For the $\a\a$ system, the same procedure is employed.
In this case, in addition to the strong potential, one has to consider also 
the presence of the long-range Coulomb interaction, $V_C$. 
Then, the $T$-matrix can be separated as follows
\begin{equation}
	T = T_{C} + T_{SC} \,,
\end{equation}
with $T_{C}$ the $T$-matrix for pure Coulomb scattering 
and $T_{SC}$ the Coulomb-distorted strong interaction part of the $T$-matrix.
The second term satisfies the following equation 
\begin{align}
	T_{SC}(\bmp,\bmp';E)
	&= \langle\psi_{\bmp}^{(-)}|V|\psi_{\bmp'}^{(+)}\rangle 
	\nonumber\\
	&+ \int\frac{d^3\bmq}{(2\pi)^3}
	\langle\psi_{\bmp}^{(-)}|V|\psi_{\bmq}^{(-)}\rangle 
	\frac{ T_{SC}(\bmq,\bmp';E) }{ E-\frac{q^2}{2\mu_{\a\a}}+i\varepsilon } 
	\,,\label{eq:LSaa}
\end{align}
where $|\psi_{\bmp}^{(\pm)}\rangle = 
\big( 1+G_C^{(\pm)}V_C \big) |\bmp\rangle$, 
$G_C^{(\pm)}$ being the Coulomb Green's function.
Then, for the $\a \a$ $S_{0}$ channel the ERE takes the form
\begin{equation}
	\frac{C_\eta^2 e^{2 i\sigma_0} }{T_{SC}^{\mathrm{on}}(E)}
	= \frac{\mu_{\a\a}}{2 \pi} \left(%
	\frac{1}{a_{0}} - \frac{1}{2} r_{0} k^{2} + O(k^{4}) + 2k_C H(\eta)
	\right)\,,\label{eq:ERE-aa}
\end{equation}
where $\eta=k_{C}/k$ with $k_C=4\a_{\mathrm{E}} \mu_{\a \a}$.
Moreover, $C_\eta^2 = 2\pi \eta /[\exp(2\pi \eta)-1]$ is the Sommerfeld factor
and $\sigma_0$ is the pure Coulomb phase shift. 
The function $H$ is related to the digamma function 
$\psi$ by $H(\eta) = \text{Re}\left[ \psi(1+i \eta)\right]- \text{ln}\eta+ \frac{i}{2\eta}C_\eta^2$~\cite{HIGA08}.
The experimental values of $a_0$ and $r_0$ employed to determine 
the $\a\a$ LECs are reported in Tab.~\ref{tab:ar_exp}.

In general, there are two different approaches to calculate the $T$-matrix,
depending on the case in question~\cite{van_Kolck_1999}.
One case arises when all the ERE parameters are of ``natural'' size, 
i.e.~given by the appropriate power of the high-momentum scale $M_{hi}$. 
This indicates a lack of low-energy resonances in the system.  
In this natural case, a perturbative approach can be used to calculate the
$T$-matrix at low scattering energy, 
by considering only the diagrams that give the largest contribution to it.
Nonetheless, in most cases of interest, a near-threshold resonance on a
momentum scale $M_{lo} \ll M_{hi}$ is present, violating the naive estimate of
the dimensional analysis. This situation can occur when one or more of the ERE
parameters have sizes related to the low-momentum scale $M_{lo}$. 
In this scenario, to describe the correct behaviour near the resonance, 
the $T$-matrix has to be resummed to all orders in the loop expansion.
In the present work, the non-perturbative approach is used, due to the presence
of low-energy resonances in the $\a n$ $P$-state and in the $\a\a$ $S$-state.
The same approach is also applied to the non-resonant $\a n$ $S$-state.
To make the truncation of the effective Lagrangian~(\ref{Lint}) meaningful, 
the power counting schemes employed for the various interactions will be
specified in the following.

For the $\a n$ system, we adopt the power counting~\cite{Bed2003,ef}
\begin{align}
	&a_1 \sim \frac{1}{M_{lo}^2 M_{hi}}\,,
	&&r_1 \sim M_{hi}\,,\label{eq:pc_anP}\\
	&a_0 \sim \frac{1}{M_{hi}}\,,
	&&r_0 \sim \frac{1}{M_{hi}}\,,\label{eq:pc_anS}
\end{align}
which supports a shallow but narrow $P$-wave resonance. 
Note that a different power counting for $P$-wave interactions was provided in
Ref.~\cite{Bert2002}, which supports a broad resonance or a bound state.
By using the ERE parameters from analysis of the scattering data 
for the $P_{3/2}$ partial wave, i.e.~$a^{\text{exp}}_{1}$ and
$r^{\text{exp}}_{1}$ from Tab.~\ref{tab:ar_exp}, we obtain
$M_{lo}\approx 50 \text{ MeV}$ and $ M_{hi}\approx 170 \text{ MeV}$; while from 
$a^{\text{exp}}_{0}$ and $r^{\text{exp}}_{0}$ (see again Tab.~\ref{tab:ar_exp}) 
we get $ M_{hi}\approx 180$ MeV as an average.
The expansion parameter is represented by the ratio 
$M_{lo,\a n} / M_{hi,\a n}\approx 0.3$, where 
$M_{hi,\a n} = \min\{ M^{S}_{hi,\a n}, M^{P}_{hi,\a n}, \}$.
Assuming~(\ref{eq:pc_anP}) and~(\ref{eq:pc_anS}), both scattering volume $a_1$
and effective range $r_1$ contribute to the Leading Order (LO).
This justifies the terms of the effective potential taken into account
in Eq.(\ref{eq:partial}).
The proper order of the $a_0$ contribution is less straightforward.
In fact, for a three-body bound-state calculation with the energy being
negative the term is only subleading~\cite{Chen2}. 
On the other hand, in the positive low-energy regime the $a_0$ contribution
should be regarded as LO~\cite{Bed2003}. 
As a consequence, in our case of low-energy photodisintegration, 
it is considered of leading order.
Note that we have included in the $\a n$ $S$-wave interaction 
also the $r_0$ term, which is of higher order.
However, we have checked that for our study of low-energy photodisintegration
there are only minor effects due to this additional higher-order correction.

For the $\a\a$ system, in addition to $M_{lo}$ and $M_{hi}$, 
also $k_C$ appears as a relevant scale of the theory, 
and we have~\cite{HIGA08,ef}
\begin{align}
	&a_0\sim\frac{M_{hi}^2}{ M_{lo}^3}\;, 
	&&r_0\sim\frac{1}{3k_C}\sim\frac{1}{ M_{hi}}\,.
\end{align}
Using here the experimental scattering parameters for the $\a\a$ $S_{0}$ partial wave, 
$a^{\text{exp}}_{0}$ and $r^{\text{exp}}_{0}$ of Tab.~\ref{tab:ar_exp}, 
we obtain approximately $M_{lo}\approx 20$ MeV and $M_{hi}\approx 170$ MeV.
This leads to $M_{lo,\a\a} / M_{hi,\a\a} \approx 0.1$.
In this case, both $a_0$ and $r_0$ give contribution to the LO.

Finally, in order to evaluate the range of validity of our EFT, 
we should also consider the breakdown scale of the $\a \a n$ system, 
which is given by 
$M_{hi} = \min\{ M_{hi,\a n}, M_{hi,\a \a} \} = 170 \text{ MeV}$.

Before looking at the low-energy phase shifts generated by each two-body
effective interaction, it is worth noting that an explicit constraint applies
to the parameter determining the range of the EFT potential, 
i.e.~the momentum cutoff $\Lambda$. 
This constraint is a consequence of the Wigner bound~\cite{Wigner}. 
For our cases, we have to fulfill the following conditions: 
$\Lambda^{S}_{\a n}<843 \text{ MeV}$, $\Lambda^{P}_{\a n}<340 \text{ MeV}$ 
and $\Lambda^{S}_{\a\a}<230 \text{ MeV}$.

Moreover, we point out that, since the procedure described at the beginning of
this section yields quadratic equations for either $\a n$ or $\a \a$ LECs, 
two sets of solutions for $\lambda_0,\lambda_1$ exist, 
one with a positive $\lambda_0$ and one with a negative $\lambda_0$.
Accordingly, in the following, these two sets will be denoted as
$\lambda_{0/1}^+$  and $\lambda_{0/1}^-$, respectively.

\begin{figure}[tbp!]
	\centering
	\includegraphics[width=\columnwidth]{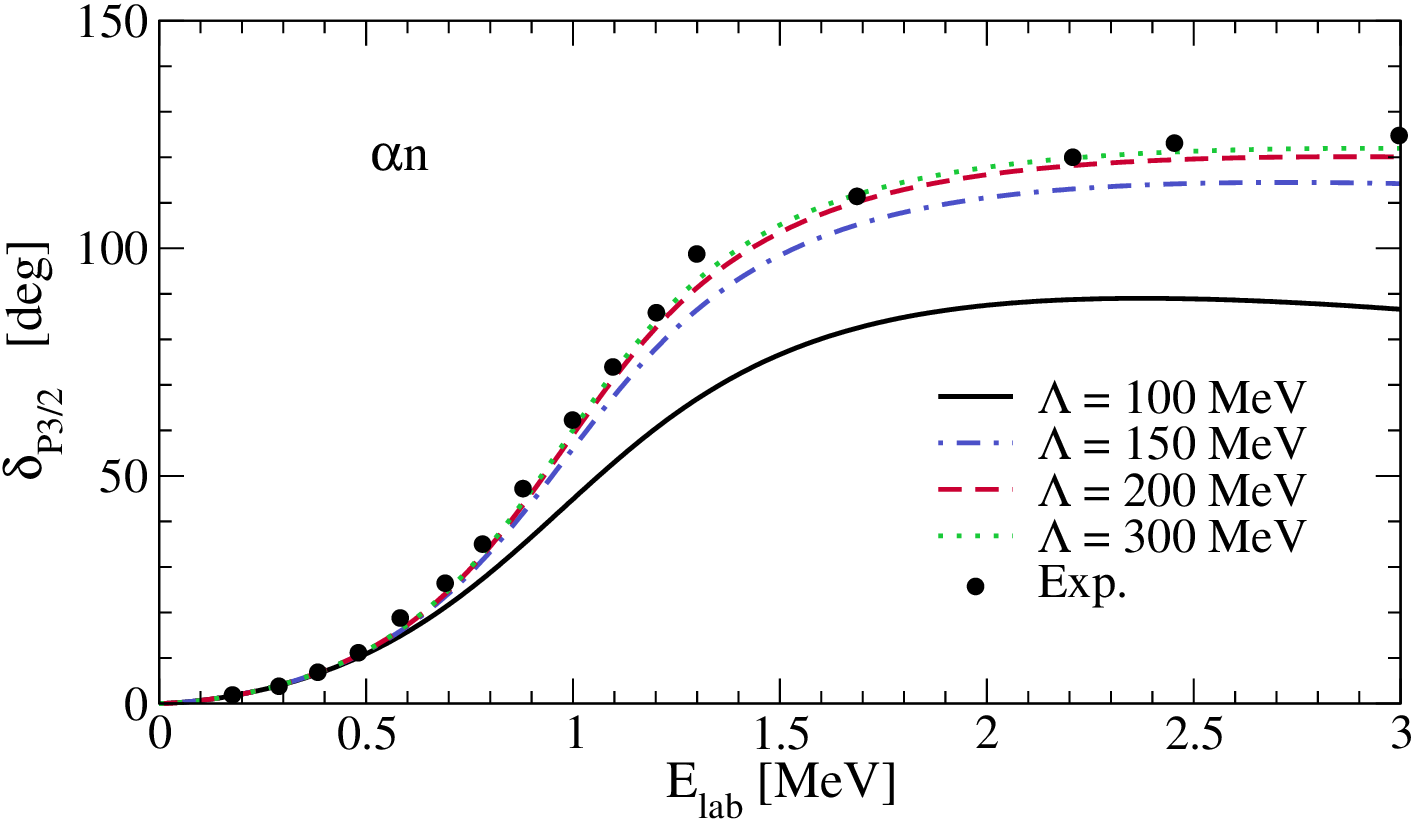}%{FIGURES/an_P32_paper}
	\caption{\label{cutoffan}%
		Calculated $\a n$ phase shifts $\delta_{P_{3/2}}$ 
		for different cutoffs $\Lambda$ as a function of the laboratory energy.
		The experimental data (circles) are taken from Ref.~\cite{morgan}.}
\end{figure}

\begin{figure}[tbp!]
	\centering
	\includegraphics[width=\columnwidth]{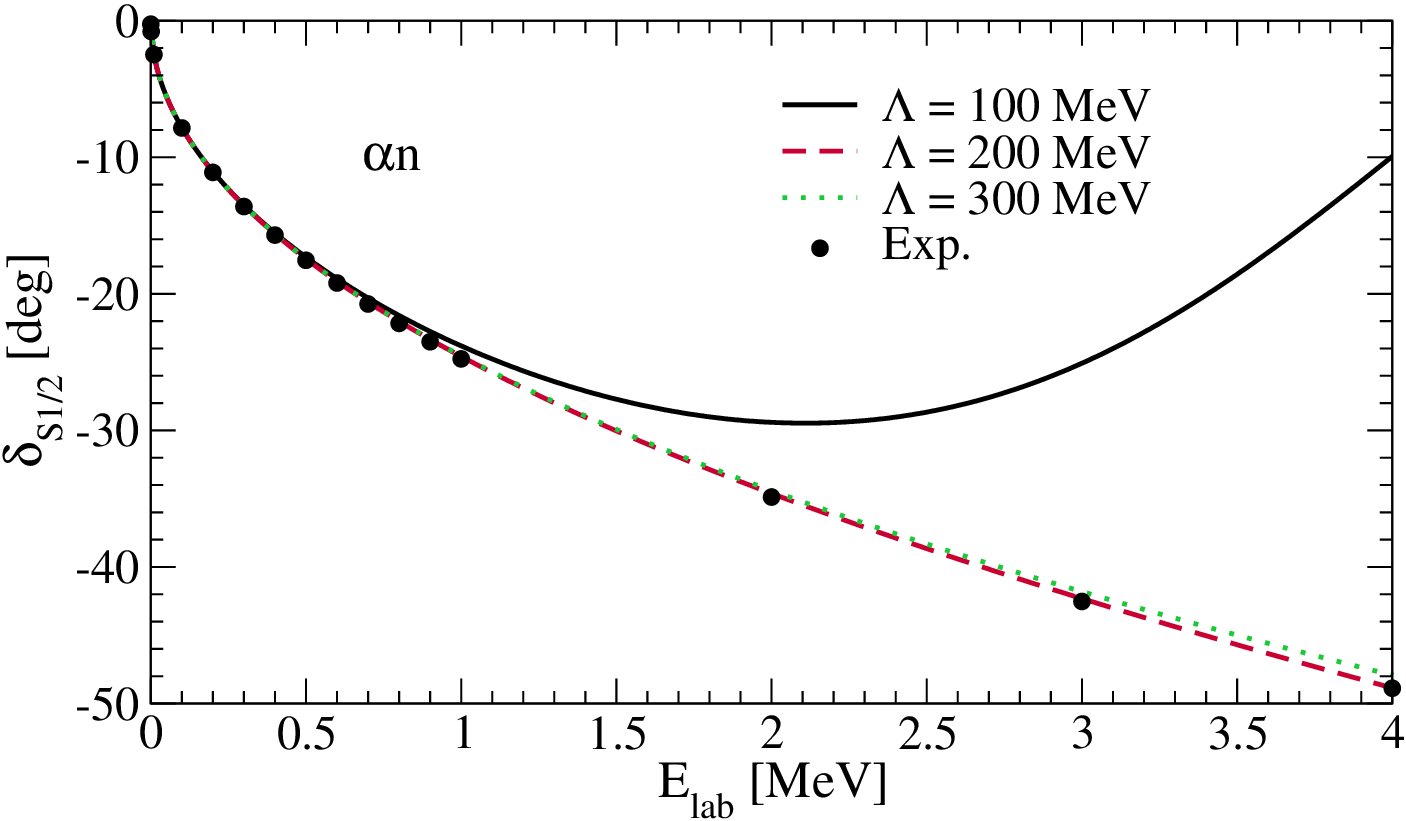}%{FIGURES/an_S12_paper}
	\caption{\label{an0}%
		Same as in Fig.~\ref{cutoffan} but for the partial wave $S_{1/2}$ 
		with results (circles) from an $R$-matrix analysis of $A=5$ 
		reaction data~\cite{Hayle}.}
\end{figure}

In Figs.~\ref{cutoffan} and~\ref{an0}%
,
we show the low-energy $\a n$ $P_{3/2}$ and $S_{1/2}$ phase shifts,
respectively, calculated with the corresponding $\lambda_{0/1}^-$ solution set.
Very similar results are obtained with the set $\lambda_{0/1}^+$. 
In both cases the momentum-regulator function $g(p)=e^{-(p/\Lambda)^{4}}$ 
has been employed.
With the higher value of the cutoff, i.e.~$300$ MeV, 
one finds a good agreement with the experimental data for both partial waves.

As a matter of fact, the  $\alpha n$ effective potential creates a deep
two-body bound state for cutoff values $\Lambda^S_{\a n} \ge 200$ MeV. 
The origin of this unphysical state lies in the absence of the Pauli exclusion
principle in the $\alpha\alpha n$  cluster approach. 
In principle, the explicit inclusion of antisymmetry effects is not required in
Cluster EFT. 
These effects are determined by wave function overlaps, which are governed by
the same expansion in powers of $M_{lo}/M_{hi}$ that underpins Cluster EFT.
However, we prefer to eliminate this spurious state to avoid introducing
unphysical states into the three-body system. To achieve this, we use the
method described in Ref.~\cite{KUKULIN1978330}, which consists in adding a
``projection'' potential to the effective $\a n$ $S$-wave interaction. 
This term is constructed by using the wave function of the deep bound state 
as follows
\begin{equation}
	V_{PR}(\bmp,\bmp') 
	\equiv \langle \bmp | \phi_S \rangle \Gamma \langle \phi_S | \bmp' \rangle
	= \Gamma \phi^*_S(\bmp') \phi_S(\bmp)\,.\label{eq:Vpr}
\end{equation}
$\Gamma$ is a parameter whose exact numerical value is not relevant, 
since formally it has to be $\Gamma \to \infty$ but in principle it is
sufficient to use a value for which the three-body bound-state and scattering
calculations turn out to be $\Gamma$-independent.

\begin{figure}[tbp!]
	\includegraphics[width=\columnwidth]{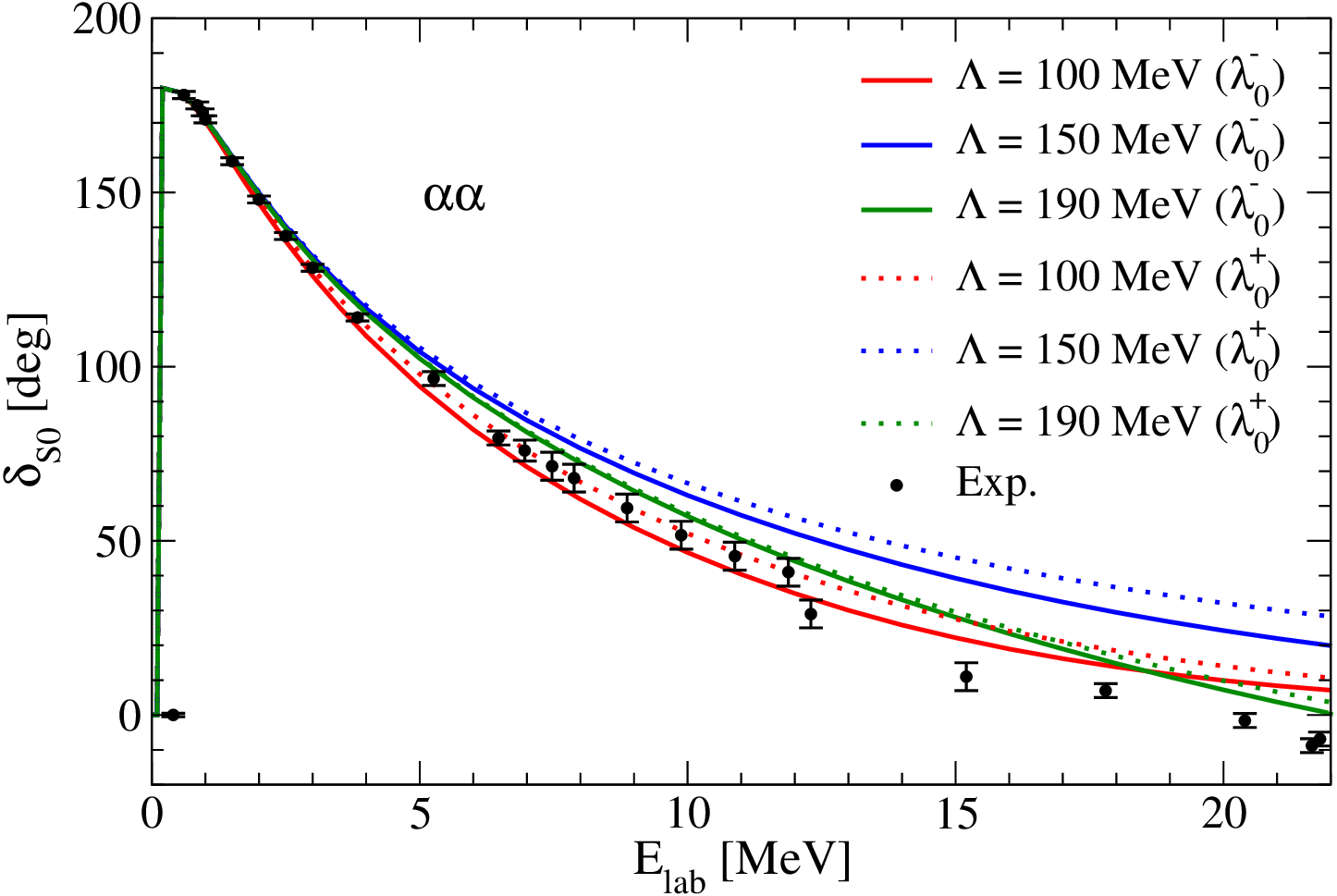}
	\caption{\label{phaseshiftaa}Calculated $\a\a$ phase shifts $\delta_{S_0}$
	for different cutoffs $\Lambda$ and for both 
	plus- and minus-set of solutions for the LECs $\lambda_0$ and $\lambda_1$
	in comparison with the experimental data (circles) from Ref.~\cite{AFZAL}.}
\end{figure}

Turning to the $\a\a$ case, in Fig.~\ref{phaseshiftaa}%
we show the $\a\a$ $S$-wave phase shifts calculated for various cutoffs with
both sets of solutions $\lambda_{0/1}^+$ and $\lambda_{0/1}^-$. 
In this case the regulator $g(p)=e^{-(p/\Lambda)^{2}}$ has been used.
It is remarkable that for the largest value of $\Lambda$ the agreement with
the experimental data is quite good even at rather high energies.

In the following we choose to use the set of coupling constants
of more natural size, namely $\lambda_{0/1}^-$.
In support of this choice, it is not difficult to demonstrate that 
only the minus-solution of the LEC $\lambda_1$ admits the value
$\lambda_1^-=0$, thus representing a continuation of the theory 
in which the effective interaction would be parametrized by only one LEC,
$\lambda_0$~\cite{Beck:2019}.

%------------------------------------------------------------------------
\subsection{\label{subsec:3bf}Three-body interaction}
%------------------------------------------------------------------------
As will be shown in Sec.~\ref{sec:GS}, the two-body potentials described here
generate a rather significant cutoff dependence of the ground-state energy of
the $\a\a n$ system. In an EFT such a dependence can be interpreted as a lack
of a three-body force. Many-body forces are generally suppressed by
power counting but, in some cases, in order to achieve a cutoff-independent
description of the observables, 
it is necessary to promote them to a lower order. 
The promotion of a three-body force to LO is a feature present in Pionless EFT
and its variants. Following these examples, we insert in our EFT a LO
three-body force given by
\begin{equation}
	\mathcal{L}^3_{\mathrm{int}} = {C_{3}} (\Psi\Psi n)^\dag (\Psi \Psi n) \,.
\end{equation}
From the expression above one obtains a regularized momentum-space potential
\begin{equation}
	V_3(Q,Q')
	=\lambda_3\,e^{-(Q/\Lambda_3)^2}e^{-(Q'/\Lambda_3)^2}\,. \label{3BF}
\end{equation}
$Q$ will be defined in Sec.~\ref{sec:GS} as the hypermomentum 
of the three-body system. 
Moreover, since $Q^2$ ($Q'^2$) is a mass-weighted combination of 
$\bmp^2_{ij}$ ($\bmp'^2_{ij}$), 
the squared relative momenta of the $ij=12,23,13$ particle pairs, 
there is a correlation between the three-body cutoff $\Lambda_3$ 
and the two-body cutoffs. 
For each choice of $\Lambda_3$, the coupling constant $\lambda_3$
is fixed on a three-body observable.

Similarly to the two-body interaction, the three-body interaction is
state-dependent. Using the same expression of $V_3$ with different coefficient
$\lambda_3$, its contribution to a specific three-body state with a given total
angular momentum $J$ is obtained through partial-wave projection of $V_3$.

%------------------------------------------------------------------------
\subsection{\label{subsec:em_current}Electromagnetic current}
%------------------------------------------------------------------------
We introduce the electromagnetic field via minimal coupling,
\begin{equation}
	\partial_{\mu} \rightarrow 
	\partial_{\mu}+ i e\hat{Q} A_{\mu},\label{minimalc}
\end{equation}
where $\hat{Q}$ is the charge operator, its action on a particle field yields
the electric charge of the particle. Furthermore, $A^\mu$ is the field of a
photon with energy $\omega_{\bmq}=|\bmq|=q$.
In the following we calculate the leading order contribution 
from the free $\a$-particle Lagrangian~\cite{Filandri}. 
With the substitution~(\ref{minimalc}) one obtains
\begin{equation}
	\mathcal{L}_{\a}^{e m} \simeq 
	-2e \Psi^\dag \Psi A_0 + \frac{i e}{m_\a} \Psi^\dag
	\Big(\overset{\rightarrow}{\nabla}-\overset{\leftarrow}{\nabla}\Big) 
	\Psi \cdot \vec{A} \,,\label{aaem}
\end{equation}
where we have considered only the linear terms in the electromagnetic field. 
The nuclear charge density and convection current of the $\a$-particle derived
from~(\ref{aaem}) are given by
%\begin{equation}
%	\rho^{[1]}_{\a_i} (p_{\mu,\a_i},q_\mu)
%	=- i e \,\delta(p_{\mu,\a_i}+q_\mu-p'_{\mu,\a_i})\,,\label{rhoNR}
%\end{equation}
\begin{align}
	\rho^{[1]}_{\a_i} (p_{\mu,\a_i},q_\mu)
	&=- i e \,\delta(p_{\mu,\a_i}+q_\mu-p'_{\mu,\a_i})\,,\label{rhoNR}\\
	\bm{J}^{[1]}_{\a_i}(p_{\mu,\a_i},q_\mu)
	&=\frac{ -i e}{2 m_{\a}} \left(2 \bmp_{\a_i}+\bmq\right)
	\delta(p_{\mu,\a_i}+q_\mu-p'_{\mu,\a_i})\,,\label{Ji}
\end{align}
with $q_\mu=(\omega_\bmq,\bmq)$ and
$p_{\mu,\a_i}=(\epsilon_{\a_i},\bmp_{\a_i})$ 
denoting four-momenta of the photon and the $\a$-particle, respectively.

In case of real photons, for which $\omega_\bmq = |\bmq|$,
only the transverse part of the current needs to be considered, hence one has
\begin{align}
	J^{[1]}_{\lambda,\a_i}(\bmp_{\a_i},\bmq)
	&= \sqrt{\frac{4\pi}{3}} \frac{-ie}{m_\a} p_{\a_i}
	Y_{1\lambda}(\uvec{p}_{\a_i})%(\theta_{\a_i},\phi_{\a_i})
	\nonumber\\
	&\times \delta(\bmp_{\mu,\a_i}+\bmq_\mu-\bmp'_{\mu,\a_i})\,,\label{JT} 
\end{align}
with $p_{\a_i}=|\bmp_{\a_i}|$ and $\lambda=\pm1$ 
corresponding to the interaction with circular polarized photons.
The superscript [1] implies that the operator is of one-body nature. 
Actually the operator in (\ref{JT}) corresponds to the usual convection current
of a particle with electric charge.
The low-energy limit is obtained in momentum space by taking $q=0$~\cite{FriarSiegert}. 
Accordingly, we take the one-body convection current operator as
\begin{equation}
	J^{[1]}_\lambda = f_1 O_{1,\lambda} \,,
\end{equation}
with the definitions
\begin{align}
	f_1 &=  -i \sqrt{\frac{4\pi}{3}} \,,\\
	O_{1,\lambda} &= \frac{1}{m_\a} \sum_{i=1}^2 p_{\a_i}
	Y_{1\lambda}(\uvec{p}_{\a_i}) \equiv \jmath_\lambda \,.\label{eq:j}
\end{align}
Note that in this limit the dependence on $\omega_\bmq$ vanishes.

Strictly speaking, additional electromagnetic interactions can arise from the
minimal substitution of derivatives in the interacting Lagrangian~\eqref{Lint},
resulting in the two-body current $J^{[2]}$, i.e.~a photon coupling to the
$\a\a$ and $\a n$ contact terms.
However, as it is explained in the following, 
the use of the Siegert theorem can avoid their actual determination.

By neglecting magnetic contributions and assuming for the photon momentum 
$\bmq = q \uvec{z}$, 
the multipole expansion of the nuclear current leads to the expression
\begin{equation}
	J_\lambda(q) = -\sqrt{2\pi} \sum_J \sqrt{2J+1} \, T^{el}_{J\lambda}(q)\,,
\end{equation}
with $\lambda=\pm1$, where $T^{el}_{J\lambda}(q)$ is the transverse electric
multipole operator. As already said, since we are studying the $^9$Be
photodisintegration reaction in the low-energy regime of astrophysical
relevance, we are allowed to apply the Siegert theorem in the long-wavelength
approximation, for which the main contribution to the electric $J=1$ transition
is given by the dipole operator. Specifically, we have
\begin{equation}
	J_\lambda = f_2(\omega_\bmq) O_{2,\lambda} \,,
\end{equation}
with
\begin{align}
	f_2(\omega_\bmq) &=  -i \omega_\bmq Z_\alpha \sqrt{\frac{4\pi}{3}} \,,\\
	O_{2,\lambda} &= \sum_{i=1}^2 r_{\a_i} Y_{1\lambda}(\uvec{r}_{\a_i}) 
	\equiv d_\lambda \,,\label{eq:d}
\end{align}
where $Z_{\a}=2$ and $\bmr_{\a_i}=r_{\a_i} \uvec{r}_{\a_i}$ is the position
vector of each charged particle in the center-of-mass reference frame.
There is an advantage in using the Siegert theorem: 
since the continuity equation is used explicitly, 
the contributions of the one-body, the two-body and the many-body current
operators, possibly originated by the potential, 
are automatically included in the calculation.

%======================================================================== GS
\section{\label{sec:GS}Calculation of the ground state}
%========================================================================
The $\Be$ ground-state wave function $\psi_0$ is calculated by solving the
Schr$\mathrm{\ddot{o}}$dinger equation via an expansion of the wave function
on the HH basis. Then one uses the Rayleigh-Ritz variational principle,
$\delta\langle\psi|H-E|\psi\rangle=0$, to find the expansion coefficients.
Since the  potentials are born in momentum space, 
we choose to work with the HH basis in this space.
We introduce the mass-weighted Jacobi momenta for the three-body system as
\begin{align}
	&\bm{\pi}_2 = \sqrt{\frac{m_RM_2}{m_1m_2}} 
	\left( \frac{m_1\bmp_2-m_2\bmp_1}{M_2} \right) \,,\label{eq:pi2}\\
	&\bm{\pi}_1 = \sqrt{\frac{m_RM_2}{m_3M_3}} 
	\left( \bmp_3-\frac{m_3}{M_2}(\bmp_1+\bmp_2) \right) \,,\label{eq:pi1}\\
	&\bm{\pi}_0 = \sqrt{\frac{m_R}{M_3}}(\bmp_1+\bmp_2+\bmp_3) \,,\label{eq:pi0}
\end{align}
with a reference mass $m_R$ and the definition $M_i=\sum_{j=1}^im_j$. 
The hyperspherical coordinates relative to the internal Jacobi vectors
$\bm{\pi}_i=\pi_i \uvec{\pi}_i$, with $i=1,2$, are given by
\begin{align}
	\pi_2 &= Q\sin\varphi_2\,,\\
	\pi_1 &= Q\cos\varphi_2\,,\label{pi}
\end{align}
where $Q$ is the hypermomentum and $\varphi_2$ the three-body hyperangle.
Here the basis functions $\Phi_{m,[K]}$ are defined as a product of a HH part,
consisting of a ``hypermomental'' function times a hyperangular function,
with the neutron spin wave function,
\begin{subequations}\label{wavefunction}
\begin{align}
	\Phi_{m,[K]}(\bm{\pi}_1,\bm{\pi}_2)
	&=\Phi_{m,[K]}(Q,\Omega^{(Q)})\otimes\ket{\chi(s_n)}
	\,,\label{wavefunction_a}\\
	&=f_m(Q)\,\mathcal{Y}_{[K]}(\Omega^{(Q)})\otimes\ket{\chi(s_n)}
	\,,\label{wavefunction_b}
\end{align}
\end{subequations}
where $\chi(s_n)$ denotes the neutron spin state. 
As hypermomental functions we choose 
\begin{equation}
	f_m(Q)=\frac{1}{\beta^3}\sqrt{\frac{m!}{(m+5)!}}
	e^{-\frac{Q}{2\beta}}{\cal L}_{m}^{(5)}\left(\frac{Q}{\beta}\right)\,,
\end{equation}
where $\beta$ is a variational parameter and ${\cal L}_{m}^{(5)}$ 
are the generalized Laguerre polynomials of order $5$.
In Eq.~(\ref{wavefunction_b}),  $\mathcal{Y}_{[K]}(\Omega^{(Q)})$ 
denotes the hyperspherical harmonics with $\Omega^{(Q)}\equiv(\uvec{\pi}_1,\uvec{\pi}_2,\varphi_2)$ 
and the quantum numbers set $[K] \equiv (\ell_{1},\ell_{2},L,M,K)$, 
where $\ell_i$ is the orbital momentum associated to the $i$-th Jacobi vector,
$L$ and $M$ are the total orbital momentum and its projection, respectively,
and $K$ is the hyperangular momentum.

The  HH functions do not possess any permutational symmetry. 
The proper symmetry between the two $\a$-particles is ensured here
by using the NSHH approach of Ref.~\cite{Deflorian}.

\begin{figure}[tbp!]
	\centering
	\includegraphics[width=\columnwidth]{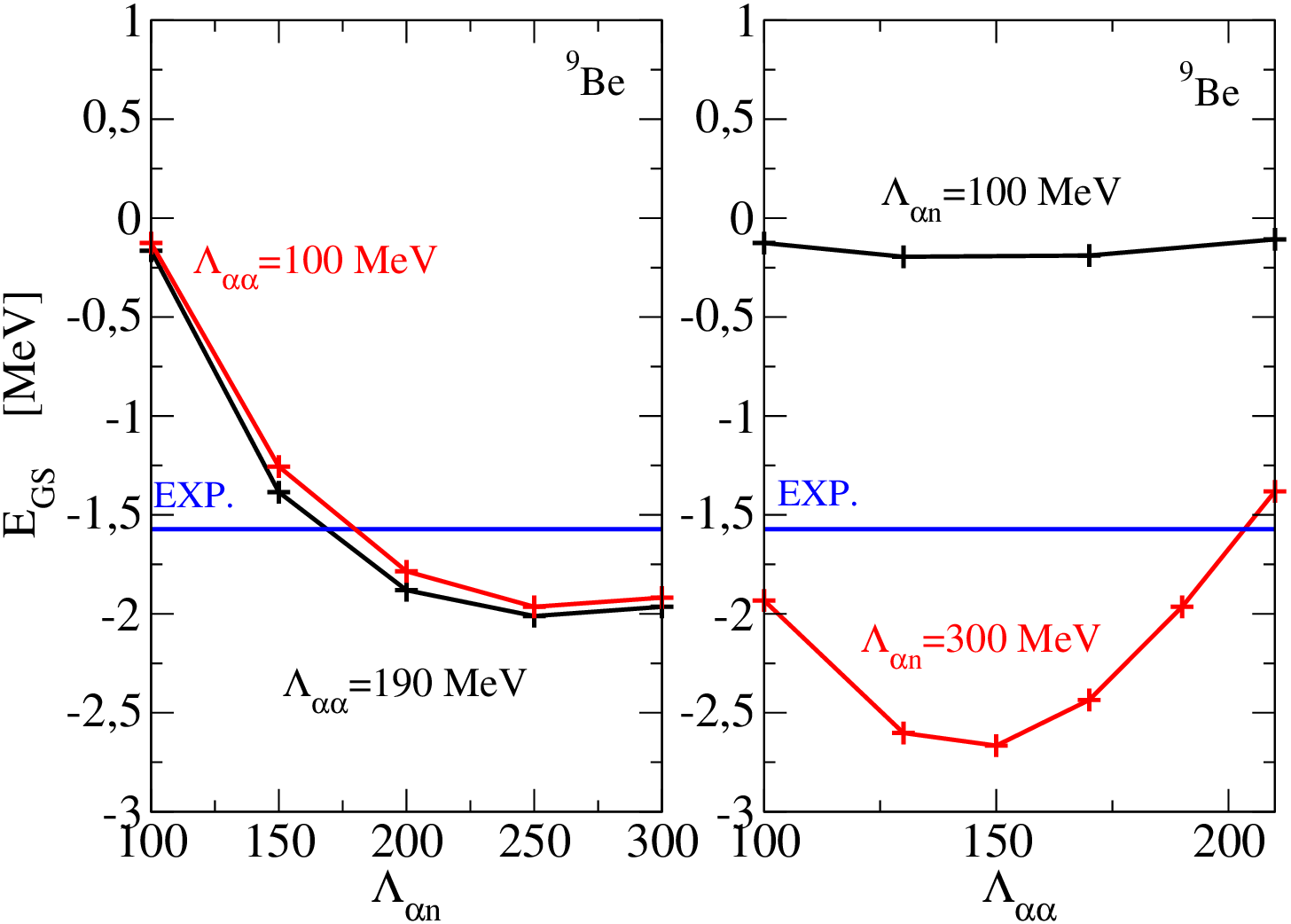}
	\caption{\label{cutoffbe}%
		(Color online) Cutoff dependence of $\Be$ ground-state energy 
		with HH parameters $\beta=0.05$ fm$^{-1}$, 
		maximum number of hypermomental functions $m_{\rm max}=30$ 
		and maximum hyperangular momentum $K_{\rm max}=23$. In the left panel
		the dependence on $\Lambda^P_{\a n}$ with $\Lambda^S_{\a \a}$ equal to
		$190$ MeV (black curve) and $100$ MeV (red curve) is shown. 
		In the right panel the dependence on $\Lambda^S_{\a \a}$ 
		with $\Lambda^P_{\a n}$ equal to $100$ MeV (black curve) and $300$ MeV
		(red curve) is visible.
		The horizontal blue line is the experimental three-body ground-state
		energy $-1.573$ MeV~\cite{TILLEY2004155}.}
\end{figure}

For the $\a\a$ and $\a n$ pairs we take the potentials introduced in the
previous section and, in addition, 
we include the Coulomb interaction for the $\a\a$ pair.
In Fig.~\ref{cutoffbe}%
the resulting cutoff dependence of the  ground-state energy can be seen.
By fixing $\Lambda^S_{\a \a}$ to $190$ MeV and varying $\Lambda^P_{\a n}$ 
from $100$ to $300$ MeV the energy varies from $-0.17$ to $-1.96$ MeV.
A similar behaviour is also obtained for $\Lambda^S_{\a \a}=100$ MeV.
It is quite a large variation, 
but all the considered cutoffs are theoretically allowed, 
since they are all smaller than the corresponding Wigner bound of 340 MeV.
The situation is different for the $\a\a$ cutoff. 
By an increase of $\Lambda^S_{\a\a}$ from $100$ to $200$ MeV, 
when fixing $\Lambda^P_{\a n}$ to $300$ MeV, 
the energy reaches a minimum of $-2.66$ MeV  at $150$ MeV and 
crosses the experimental line close to the Wigner bound at $230$ MeV.
For $\Lambda^P_{\a n} = 100$ MeV one gets a rather small binding energy
with almost no $\Lambda^S_{\a\a}$ dependence.

The stronger cutoff dependence on $\Lambda_{\a n}^P$ could be interpreted 
as a residue of a discrete scale invariance~\cite{kievsky} 
for three-body systems involving $P$-wave pairwise interactions. 
The discrete scaling characteristic of the Efimov effect~\cite{EFIMOV} 
can occur, in fact, not only for three identical bosons with $S$-wave pairwise
interactions but also trying to describe two pairwisely non-interacting
identical particles each having a $P$-wave resonance with a third
particle~\cite{Pwaveefimov}. 
Although the exact discrete scaling invariance in the $P$-wave unitary limit
is unphysical~\cite{Chenhalo,Chen2,nopwaveef}, it has been shown in Cluster EFT
study of $^6$He as a $\alpha nn$ system with two pairwise $P$-wave $\a n$
interactions that a discrete scaling with a varying scaling factor is observed
when varying cutoff in this system~\cite{Ji:2014wta,Li:2023hwe}.
On the contrary, the weaker cutoff dependence on $\Lambda_{\alpha\alpha}^S$
can be understood by the fact that Coulomb barrier at short distance
prevents particles interacting at shorter distance to form a deeper states.

\begin{figure}[tbp!]
	\centering
	\includegraphics[width=\columnwidth]{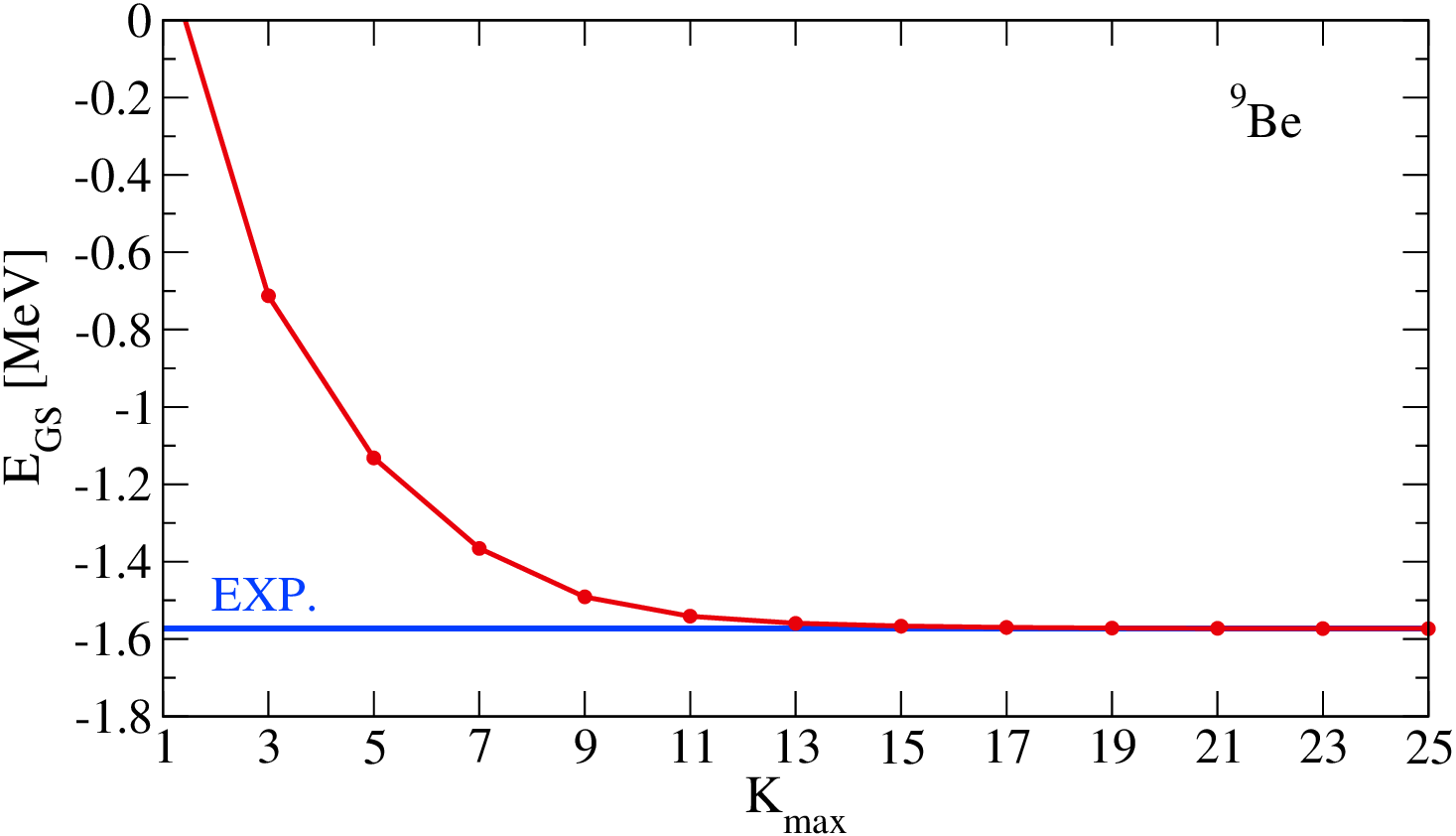}
	\caption{\label{bektb1}%
		(Color online) Convergence study of the $\Be$ ground-state energy 
		with respect to the maximum hyperangular momentum $K_{\rm max}$ 
		for a Hamiltonian that includes the three-body force of Eq.~(\ref{3BF})
		with $\Lambda_3=300$ MeV. The chosen two-body cutoffs are
		$\Lambda^S_{\a\a}=190$ MeV and $\Lambda^P_{\a n}=300$ MeV.
		The experimental three-body ground-state energy (blue line)
		is $-1.573$ MeV as in Ref.~\cite{TILLEY2004155}.}
\end{figure}

The cutoff dependence of the ground-state energy is eliminated 
by adding the three-body force of Eq.~(\ref{3BF}). In fact, 
as already mentioned, the three-body force contributes to lowest order.
Then, for a specific parameter setting of the two-body potentials 
and any choice of the cutoff $\Lambda_3$ of the three-body force, 
we fit the coupling constant $\lambda^{3/2^-}_3$ 
to reproduce the experimental $\Be$ ground-state energy.
In Fig.~\ref{bektb1}%
we show a typical pattern of the HH convergence of the ground-state energy.
It is evident that a sufficiently convergent result is reached already for a rather low $K_{\rm max}$.

%==========================================================RESULTS-dipole
\section{\label{sec:cross}Low-energy photodisintegration cross section}
%========================================================================
%------------------------------------------------------------------------
\subsection{\label{subsec:lit_calc}Calculation of the integral transform}
%------------------------------------------------------------------------
From a practical point of view, the integral transform of Eq.~(\ref{2.13})
is evaluated by expanding the LIT function $\tilde{\Psi}_i$ 
on a basis of localized functions of dimension $\nu_{\text{max}}$.
This leads to the following expression~\cite{Efros_2007}
\begin{equation}
	L_i(\sigma_R,\sigma_I) 
	= \sum_{\nu=1}^{\nu_{\text{max}}} 
	\frac{ | \langle \psi_{\nu} | \hat{O}_{i,\lambda} | \psi_0 \rangle |^2  }%
	{ (E_{\nu}-E_0-\sigma_R)^2+\sigma_I^2 } \,,\label{eq:L_eigenv_method}
\end{equation}
with $\hat{H} |\psi_{\nu}\rangle = E_{\nu} |\psi_{\nu}\rangle$ and $i=1,2$.
The operators $\hat{O}_{1,\lambda}$ and $\hat{O}_{2,\lambda}$ are the 
one-body current $\hat{\jmath}_\lambda$ and the dipole $\hat{d}_\lambda$ 
as defined in Eqs.~(\ref{eq:j}) and~(\ref{eq:d}), respectively.
In order to reduce the computational time, we also use the Lanczos method
to calculate $L_i(\sigma_R,\sigma_I)$~\cite{Marchisio}.

When $i=1$, the matrix elements in Eq.~(\ref{eq:L_eigenv_method}) 
are evaluated by using the HH basis defined in the previous section:
\begin{align}
	{\langle \hat{\jmath}_{\lambda} \rangle}_{\nu 0} &=
	\int d\Omega^{(Q)} dQ \, Q^5 \nonumber\\
	&\times \psi^*_{\nu}(Q,\Omega^{(Q)}) 
	\;\jmath_\lambda(Q,\Omega^{(Q)})\; 
	\psi_0(Q,\Omega^{(Q)}) \,,\label{eq:j_me}
\end{align}
with $\psi_{\nu/0}(Q,\Omega^{(Q)})
=\sum_{m [K]} c^{\nu/0}_{m,[K]} \Phi_{m,[K]} (Q,\Omega^{(Q)})$.
Since the correct symmetry of the wave function is taken into account 
by the NSHH approach, we are free to choose the particle ordering 
in the definition of the Jacobi momenta. 
Taking the neutron as the third particle, the operator does not depend on the
Jacobi momentum $\bm{\pi}_2$, and explicitly we have~\cite{Filandri}
\begin{equation}
	\jmath_\lambda(Q,\Omega^{(Q)}) 
	= -\sqrt{ \frac{ 2m_\a }{ (2m_\a +m_n) } } 
	\frac{ Q \cos\varphi_2 }{ m_\alpha } Y_{1\lambda}(\uvec{\pi}_1)
	\,.\label{otimespi1}
\end{equation}

In the $i=2$ case, since the Siegert operator has been defined in coordinate
space, we choose to calculate the matrix elements 
${\langle \hat{d}_{\lambda} \rangle}_{\nu 0} $ in the same space. 
As already stated in Sec.~\ref{sec:GS}, the basis on which we diagonalize the
Hamiltonian is defined in momentum space. As a consequence, the proper basis
functions to use in coordinate space have to be constructed in a consistent
way. This is achieved by using the intrinsic Jacobi coordinates
\begin{align}
	&\bm{\eta}_2 
	= \sqrt{ \frac{m_1m_2}{m_RM_2} } \left( \bmr_2-\bmr_1 \right) \,,\\
	&\bm{\eta}_1 = \sqrt{ \frac{M_2m_3}{m_RM_3} } \left( \bmr_3-\frac{m_1\bmr_1+m_2\bmr_2}{M_2} \right) \,,
\end{align}
where $\bmr_i$ are the position vectors of the three particles of masses $m_i$,
with $i=1,2,3$, and $M_2$,$M_3,m_R$ are defined as in Sec.~\ref{sec:GS}.
In fact, these coordinates represent the conjugate variables of
$\bm{\pi}_2,\bm{\pi}_1$.
Each modulus $\eta_i$ is given in terms of the hyperradius $\rho$ 
and the hyperangle $\varphi_2$ as
\begin{align}
	\eta_2 &= \rho \sin\varphi_2\,,\label{eq:eta2}\\
	\eta_1 &= \rho \cos\varphi_2\,,\label{eq:eta1} 
\end{align}
leading to the following definition 
for the basis $\Phi_{m,[K]}$ in coordinate space:
\begin{equation}
	\Phi_{m,[K]}(\bm{\eta}_1,\bm{\eta}_2)
	=g_{mK}(\rho)\,\mathcal{Y}_{[K]}(\Omega^{(\rho)})\otimes\ket{\chi(s_n)}
	\,.\label{wavefunction_rspace}
\end{equation} 
We point out that, with respect to the momentum-space basis defined in
Eqs.~(\ref{wavefunction}), the hyperangular functions are unchanged in form
$\mathcal{Y}_{[K]}(\Omega^{(Q)}) \to \mathcal{Y}_{[K]}(\Omega^{(\rho)})$
[$\Omega^{(\rho)}\equiv(\uvec{\eta}_1,\uvec{\eta}_2,\varphi_2)$],
while the hyperradial term is constructed from the functions $f_m(Q)$
as~\cite{Viviani2006}
\begin{equation}
	g_{mK}(\rho) = i^K \int dQ \frac{ Q^{5} }{ (Q\rho)^{2} } 
	J_{K+2}(Q\rho) f_m(Q) \,,\label{eq:gmk(rho)}
\end{equation}
$J_{K+2}(Q\rho)$ being a Bessel function.
Thus the matrix elements in Eq.~(\ref{eq:L_eigenv_method}) 
are evaluated as~\cite{Capitani}
\begin{align}
	{\langle \hat{d}_{\lambda} \rangle}_{\nu 0} &=
	\int d\Omega^{(\rho)} d\rho \, \rho^5 \nonumber\\
	&\times \psi^*_{\nu}(\rho,\Omega^{(\rho)})
	\;d_\lambda(\rho,\Omega^{(\rho)})\; \psi_0(\rho,\Omega^{(\rho)})
	\,,\label{eq:d_me}
\end{align}
with $\psi_{\nu/0}(\rho,\Omega^{(\rho)})
=\sum_{m [K]} c^{\nu/0}_{m,[K]} \Phi_{m,[K]} (\rho,\Omega^{(\rho)})$ and
\begin{equation}
	d_\lambda(\rho,\Omega^{(\rho)})
	=   \frac{-2 m_n}{\sqrt{2m_\alpha(2m_\a+m_n)}} \rho \cos\varphi_2 \,
	Y_{1\lambda}(\uvec{\eta}_1) \,,	\label{eq:D_lambda_eta_12aa}
\end{equation}
having assumed the particle ordering $123=\a\a n$.

Since the Hamiltonian is rotationally invariant, the LIT calculation can be
carried out for a given final state $|J_fM_f\rangle$. The $\Be$ ground state
is characterized by $J_0^{\pi_0}=3/2^-$, and therefore the final states allowed
by $E1$ transitions are $J_f^{\pi_f} = 1/2^+, 3/2^+, 5/2^+$.
In the next section we will first focus on the $1/2^+$ resonance, 
being the one occurring at the lowest energy. 
Then, in order to evaluate the total photodisintegration cross section up to
photon energy $\approx 5$ MeV, 
we will take into account also the other $3/2^+$ and $5/2^+$ states.
Unless explicitly stated, the Siegert operator is used in the calculations.

%------------------------------------------------------------------------
\subsection{\label{subsec:12+}$1/2^+$ resonance}
%------------------------------------------------------------------------
As a consequence of the discussion presented in Secs.~\ref{subsec:LEC},
\ref{subsec:3bf} and~\ref{sec:GS}, for the computation of the $\Be$
ground-state wave function we choose the following HH basis parameters:
$\beta=0.05$ fm$^{-1}$, $m_{\rm max}=30$, $K_{\max}=25$.
The chosen two-body cutoffs are 
$\Lambda^S_{\a\a}=190$ MeV, $\Lambda^P_{\a n}=300$ MeV 
and $\Lambda^S_{\a n}=300$, 
being those that best reproduce the $\a n$ and $\a\a$ low-energy phase shifts
(see Figs.~\ref{cutoffan},~\ref{an0}~and~\ref{phaseshiftaa}).
The values of the LECs relative to the different two-body potentials
are listed in Tab.~\ref{LECs}.%
\begin{table}[tbp!]
	\caption{Setting of the two-body potentials parameters.}\label{LECs}
	\begin{ruledtabular}
    \begin{tabular}{ccccc}
		&$\ell_j$	&$\lambda_0\,[\textrm{fm}^{2\ell+2}]$
		&$\lambda_1\,[\textrm{fm}^{2\ell+4}]$	&$\Lambda\,[\textrm{MeV}]$\\
		\hline
		$\a\a$	&$S_0$		&$-8.03$	&5.88	& 190\\
		$\a n$	&$S_{1/2}$ 	&$-6.96$	&0.81	& 300\\
		$\a n$	&$P_{3/2}$	&$-11.35$	&4.45	& 300\\
	\end{tabular}
	\end{ruledtabular}
\end{table}
The same Hamiltonian is used to perform LIT calculations. 
However, the three-body force is state-dependent: for each cutoff $\Lambda_3$,
while for ground-state calculations the strength $\lambda^{3/2^-}_3$ is fixed
to reproduce the $\Be$ experimental three-body binding energy, 
in evaluating the LIT we choose a value of $\lambda^{1/2^+}_3$ 
for which the resonance peak is in the correct position.

\begin{figure}[tb]
	\includegraphics[width=\columnwidth]{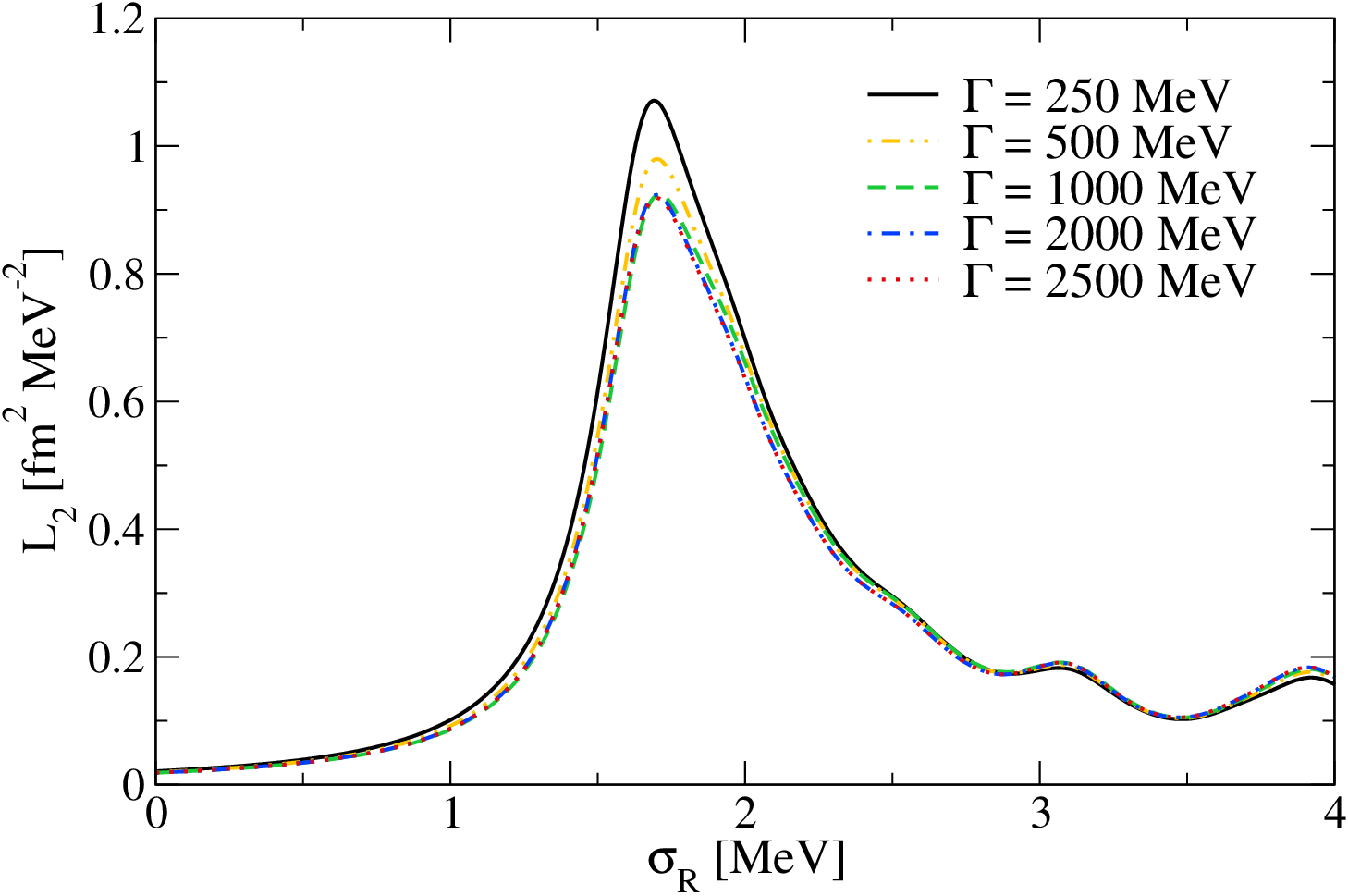}
	\caption{\label{fig:lit_s0.2_GammaG2}$1/2^+$ LIT calculated with the 
	HH basis parameters $\beta=0.05$ fm$^{-1}$, $K_\mathrm{max}=26$ and
	$m_\mathrm{max}=30$ for different values of the projection parameter
	$\Gamma$ and for $\sigma_I=0.2$ MeV. 
	The three-body cutoff is fixed at $\Lambda_3=300$ MeV.}
\end{figure}

When $\Lambda^S_{\a n} = 300$ MeV, an $\a n$ forbidden bound state 
of energy $-12.25$ MeV is present.
As the projection potential of Eq.~(\ref{eq:Vpr}) is required in the
calculation [with $\Gamma$ including the normalization factor $1/(2\pi)^3$],
we have performed some checks in the two-body 
as well as in the three-body sector~\cite{Capitani}.
The $\a n$ $\delta_{{S}_{1/2}}$ low-energy phase shifts of Fig.~\ref{an0}
are unchanged when $V_{PR}$ is used and the parameter $\Gamma$ is varied. 
Moreover, for values $\Gamma>15$ MeV 
the $\a n$ bound state is no longer present. 
Above $\Gamma \sim 750$ MeV the strength $\lambda^{3/2^-}_3$ capable of
reproducing the experimental ground-state energy for $\Lambda_3=300$ MeV
reaches a plateau. 
Furthermore, in Fig.~\ref{fig:lit_s0.2_GammaG2}%
we show the stability of the $1/2^+$ LIT by varying $\Gamma$, 
which is reached for values above $10^3$ MeV.
As a consequence, to ensure the independence of the results on the projection
parameter, we have chosen to use $\Gamma=2000$ MeV in our calculations.

\begin{figure}[tb]
	\includegraphics[width=\columnwidth]{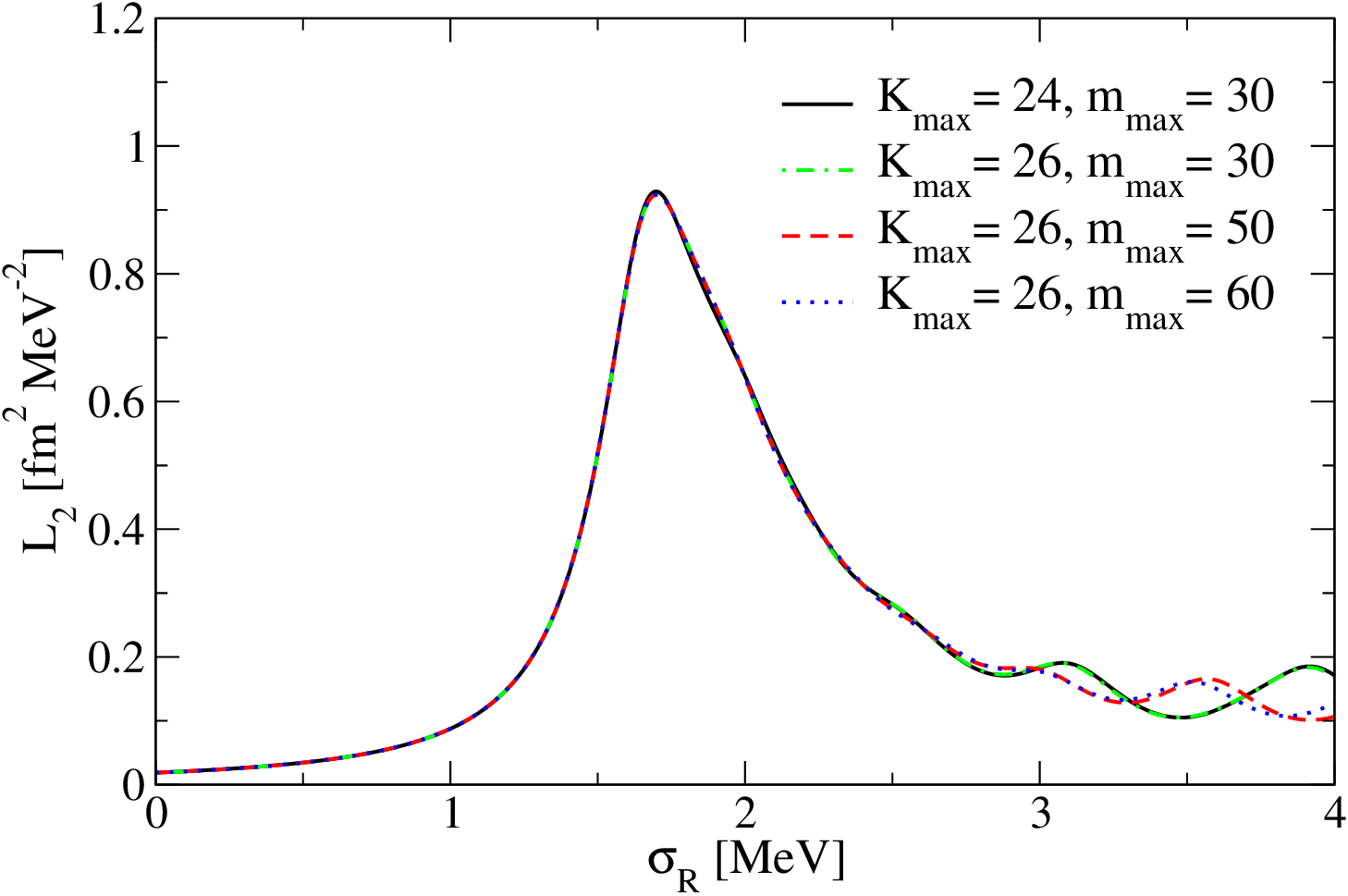}%{FIGURES/LIT/lit_s0.2_GammaG4000_R_paper}
	\caption{\label{lit_s0.2_GammaG4000_R}%
		Same as in Fig.~\ref{fig:lit_s0.2_GammaG2} but here the projection
		parameter is fixed at $\Gamma=2000$ MeV and a convergence study 
		in $K_\mathrm{max}$ and $m_\mathrm{max}$ is performed.}
\end{figure}

The computation of the integral transform $L_i(\sigma_R,\sigma_I)$ 
also requires a choice of the parameter $\sigma_I$. 
Smaller values of $\sigma_I$ ensure a better resolution of the transform, 
but generally lead to a slower convergence because the size of the basis has to
be enlarged. As done for the ground state, the set of the HH basis parameters
is chosen to give a sufficiently convergent result for the transform.
Convergence studies of the $1/2^+$ LIT computed with $\sigma_I=0.2$ MeV 
are shown in Fig.~\ref{lit_s0.2_GammaG4000_R}%
both in $K_\text{max}$ and $m_\text{max}$.
As can be deduced, the inclusion of a larger number of basis states through
$m_\text{max}$ leads to a less oscillatory behaviour in the tail beyond the
peak, also ensuring more stable results from the inversion procedure. 
The optimal values to use are therefore $K_{\text{max}}=26$ 
and $m_{\text{max}}=60$.

\begin{figure}[tb]
	\includegraphics[width=\columnwidth]{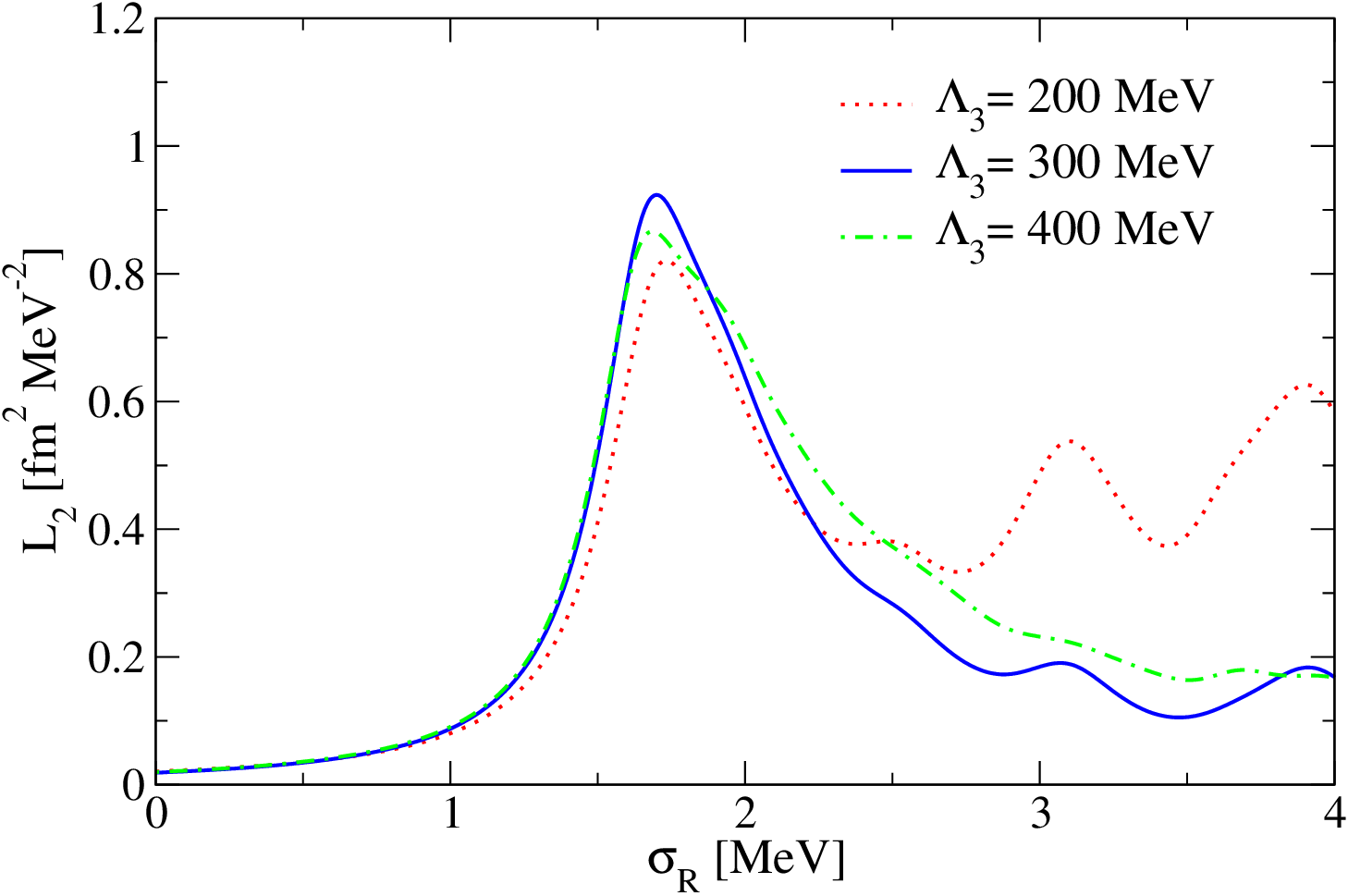}
	\caption{\label{fig:9Be_GammaG4000_tb}%
		Same as in Fig.~\ref{fig:lit_s0.2_GammaG2}, with $\Gamma=2000$ MeV. 
		The $1/2^+$ LIT is calculated here for different three-body cutoffs
		$\Lambda_3$.}
\end{figure}

We have also studied our results by varying the three-body cutoff $\Lambda_3$,
see Fig.~\ref{fig:9Be_GammaG4000_tb}.
By focussing on the main resonance peak, 
the LITs calculated with $\Lambda_3=200$ MeV and $\Lambda_3=400$ MeV are
approximately $10\%$ and $6\%$ smaller than the LIT obtained with
$\Lambda_3=300$ MeV. Thus our results show a slight dependence on $\Lambda_3$.
By excluding the $\Lambda_3=200$ MeV LIT, whose rather high and oscillatory
``background'' tail could spoil the result of the response function, 
each of the other two LITs could be used to perform the inversion.
Notice that this is consistent with the correlation relation between
$\Lambda_3$ and the chosen two-body cutoffs, 
which gives approximately $\Lambda_3 \approx 376$ MeV.

The standard inversion method used here 
consists in the following ansatz~\cite{Efros_2007}
\begin{equation}
	R\left(\epsilon\right)=\sum_{n=1}^{N} c_n 
	\chi_{n}\left(\epsilon, \alpha_{i}\right) \,.\label{RsigmR}
\end{equation}
The unknown expansion coefficients are obtained by a fit to the calculated LIT,
which can be expressed as 
\begin{equation}
	L\left(\sigma_{R}\right) = \sum_{n=1}^{N} c_n \bar{\chi}_{n}
	\left(\sigma_{R}, \alpha_{i}\right) \,,\label{LsigmR}
\end{equation}
with the definition
\begin{equation}
	\bar{\chi}_{n}\left(\sigma_{R}, \alpha_{i}\right) 
	= \int_{0}^{\infty} d \epsilon 
	\frac{\chi_{n}\left(\epsilon,\alpha_{i}\right)}%
	{\left(\epsilon-\sigma_{R}\right)^{2}+\sigma_{I}^{2}}\,. \label{chin}
\end{equation}
Here $\epsilon=\omega-\epsilon_{\mathrm{th}}$, where $\epsilon_{\mathrm{th}}$
is the threshold energy for the breakup into the continuum.
The basis set of functions $\chi_{n}$ in Eqs.~(\ref{RsigmR}) or~(\ref{chin})
is given explicitly by
\begin{align}
	\chi_1\left(\epsilon,\alpha_i\right) 
	&= \big( 1-e^{-\epsilon/{\text{MeV}}} \big)^{\alpha_1} 
	\frac{ 1 }{ (\epsilon - \alpha_2)^2 + \alpha_3^2 }  \,,\\
	\chi_n\left(\epsilon,\alpha_i\right) 
	&= \epsilon^{\alpha_4} \exp\left\{-\frac{\alpha_5\epsilon}{n-1}\right\} \,,
\end{align} 
with $n= 2, \dots, N$.
While the $n>1$ terms represent a standard basis with non-linear variational
parameters $\alpha_{4},\alpha_{5}$, here we have also included an explicit
resonant structure of Lorentzian form through the function $\chi_1$ and the
additional parameters $\alpha_1,\alpha_2,\alpha_3$. 
Since in our case narrow resonances are present, this is advantageous to avoid
the number of required functions $\chi_n$ getting too large.
By varying the non-linear parameters $\alpha_i$ on a grid, 
the linear parameters $c_{n}$ are determined for any set $\set{\alpha_i}$ 
from a least-square best fit of Eq.~(\ref{LsigmR}) 
to the calculated transform $L\left(\sigma_{R},\sigma_I\right)$. 
Finally, for each value of $N$, the best fit is selected, and the procedure is
repeated with $N \to N+1$ until a sufficient convergence is reached.

The response function obtained by inverting the integral transform in
Fig.~\ref{fig:9Be_GammaG4000_tb}%
\begin{figure}[tbp!]
	\includegraphics[width=\columnwidth]{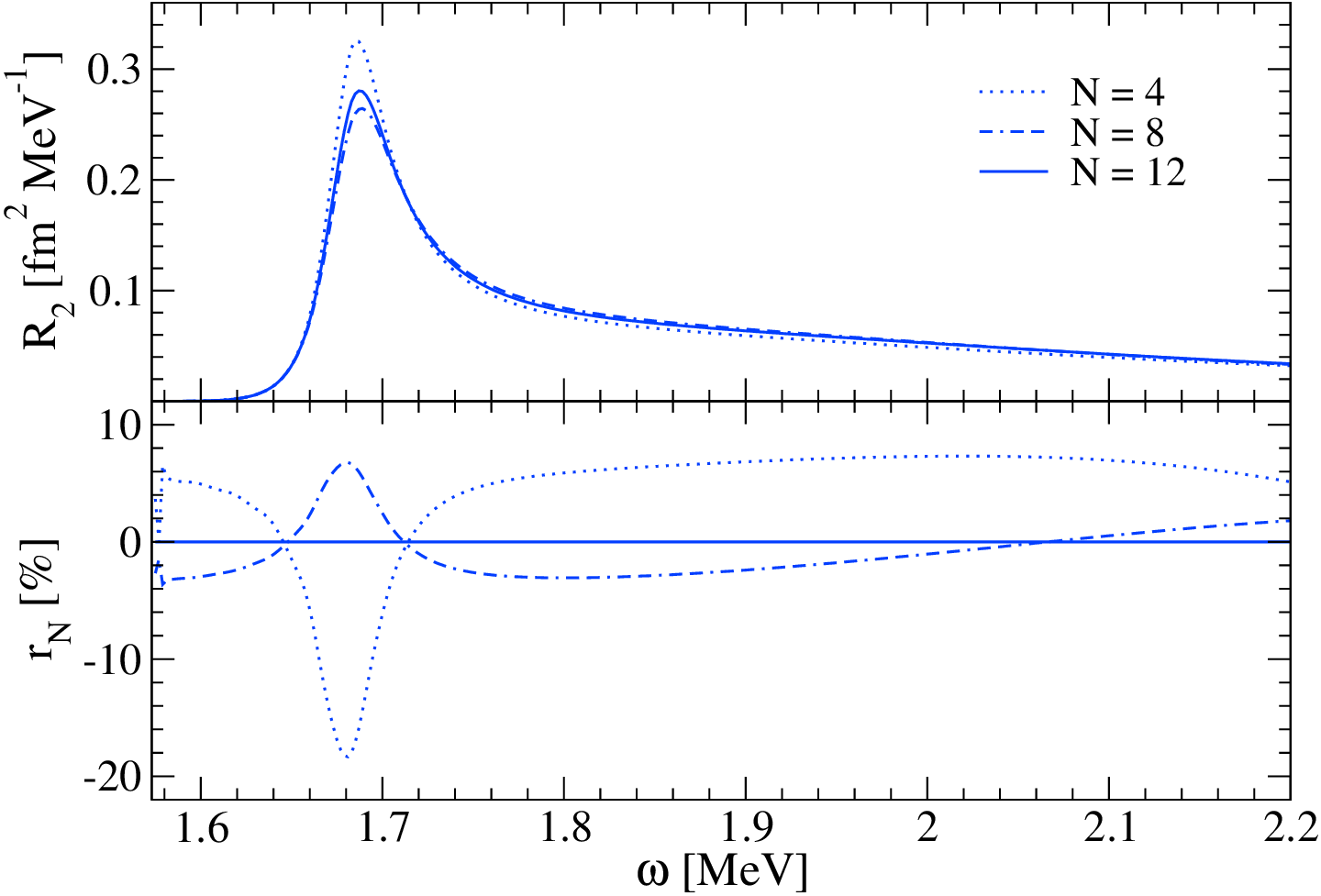}
	\caption{\label{fig:9Be_inversion_12}%
		Calculated $1/2^+$ response function for an increasing number $N$ 
		of basis functions employed in the inversion procedure (upper panel).
		The associated percentage difference $r_N$ 
		(see text for the definition) is also shown (lower panel).
	}
\end{figure}
with cutoff $\Lambda_3=300$ MeV can be seen in Fig.~\ref{fig:9Be_inversion_12},
which shows the results for an increasing number of basis functions $N$.
In the lower panel also the ratio $r_N(\omega) 
= 100 \big[R^{(\bar{N})}(\omega)-R^{(N)}(\omega)\big]/R^{(\bar{N})}(\omega)$ 
is represented, being the percentage difference of each result with respect to
the one obtained with a basis of dimension $\bar{N}=12$. 
The results are rather stable.
The major source of error lies in the lower part of the peak region,
around energy $\approx 1.68$ MeV, where a percentage $\sim 18\%$ 
is found for $N=4$, decreasing to $\sim 6\%$ for $N=8$. 
In the latter case, at most an error of $3\%$ is found in the tail. 
The difference between the response functions obtained with $N=8$ and $12$ 
can thus be taken as an estimate of the error due to the inversion procedure.

\begin{figure*}[tbp!]
	\includegraphics[width=1.25\columnwidth]{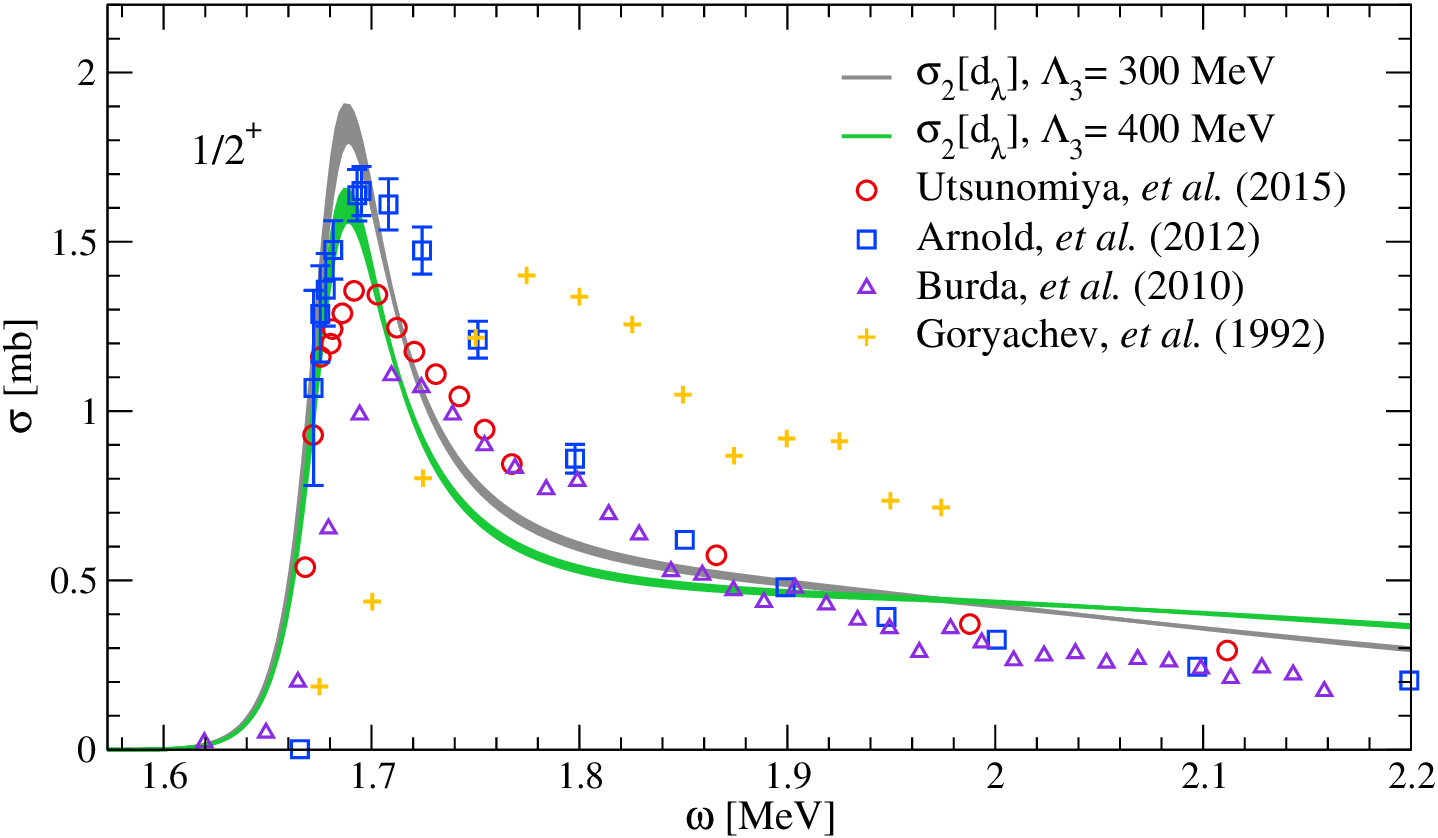}%{FIGURES/CS/9Be_cs_12_paper}
	\caption{\label{fig:9Be_cs_12}%
		(Color online) $1/2^+$ contribution to the $\Be$ photodisintegration
		cross section in comparison with different sets of experimental data
		from Refs.~\cite{utsunew,arnold,Burda,Goryachev}.
		Both $\Lambda_3=300$ MeV (grey) and $\Lambda_3=400$ MeV (green) 
		results obtained by using the dipole operator are included. 
		Lines thickness represents the assumed error 
		due to the inversion procedure. 
		The three-body threshold energy is $\epsilon_{\mathrm{th}}=1.573$ MeV.
	}
\end{figure*}

From the calculated response function $R_{2}$, we have computed the cross
section as a function of the photon energy $\omega$ by means of
Eq.~(\ref{cross}) obtaining the results in Fig.~\ref{fig:9Be_cs_12}.%
Both results obtained by inversion of the LIT calculated with 
$\Lambda_3=300$ MeV and $\Lambda_3=400$ MeV are reported, 
including also different sets of experimental data.
By inspection of the plot, we can see that our results are in fair agreement
with the data from Ref.~\cite{arnold} by Arnold, \emph{et al.}. 
Since the peak position has been fitted by adjusting the strength of the
three-body force, it is interesting to compare the calculated cross section for
peak height and peak width with experiment. 
In the $\Lambda_3=300$ MeV case, $1.85$ mb can be taken as the maximum value of
$\sigma$, with an estimated error $\pm0.05$ mb from the inversion procedure.
Using the cutoff $\Lambda_3=400$ MeV, the peak is lower in height
($1.60\pm0.05$ mb), showing a better agreement with the experiment.
The widths are small if compared with the distribution of the experimental data
points coming from the different measurements. However, above $\approx2$ MeV
the calculated cross sections overestimate all the experimental data.

%------------------------------------------------------------------------
\subsection{\label{subsec:total}The total cross section}
%------------------------------------------------------------------------
To compare our theoretical results with the experimental low-energy
photodisintegration cross section up to $5$ MeV, the contributions due to the
$E1$ transitions to the final states $5/2^+$ and $3/2^+$ must be included,
whose associated resonances occur at $\approx 3$ MeV and 
$\approx 4.7$ MeV~\cite{TILLEY2004155}, respectively.

\begin{figure}[tbp!]
	\includegraphics[width=\columnwidth]{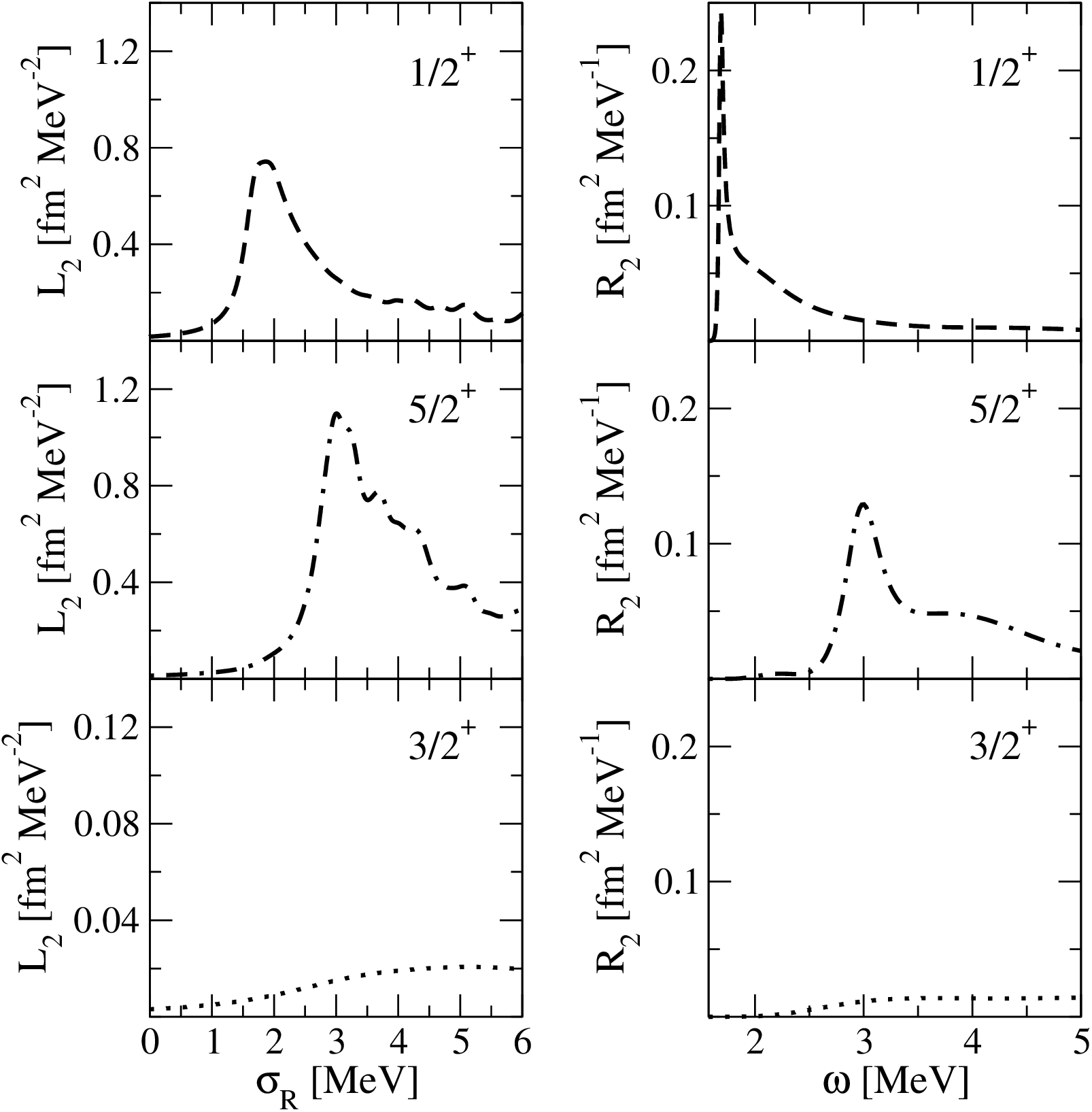}
	\caption{\label{fig:9Be_inversion_all}%
		$1/2^+$, $5/2^+$ and $3/2^+$ LITs calculated with $\sigma_I=0.2$ MeV 
		in the first two cases and $\sigma_I=1.4$ MeV 
		in the last case (left panels). 
		The three-body cutoff is $\Lambda_3=400$ MeV.
		The corresponding response functions obtained 
		from an inversion procedure are also shown as a function 
		of the photon energy $\omega$ (right panels).
	}
\end{figure}

In Fig.~\ref{fig:9Be_inversion_all}%
the $5/2^+$ and $3/2^+$ integral transforms are collected together 
with the associated response functions.
The HH basis parameters used to compute the $5/2^+$ LIT are 
$K_{\text{max}}=34$ and $m_{\text{max}}=80$, 
and those used in the $3/2^+$ case are 
$K_{\text{max}}=26$ and $m_{\text{max}}=30$. 
Being the $3/2^+$ resonance broader, calculations converge faster then in the
$5/2^+$ case. The latter requires higher values of both maximum hyperangular
momentum and number of hypermomental (hyperradial) functions, even with respect
to the $1/2^+$ channel. For the same reason, the adopted value of $\sigma_I$ is
greater in the $3/2^+$ case, $\sigma_I=1.4$ MeV; in the $5/2^+$ LIT the value
$\sigma_I=0.2$ MeV is used, just as in $1/2^+$ calculations.
Concerning the three-body interaction, the chosen cutoff is 
$\Lambda_3=400$ MeV, while the strengths $\lambda^{5/2^+}_3$ and
$\lambda^{3/2^+}_3$ are tuned to have each resonance located in the correct
position.
For completeness, in Fig.~\ref{fig:9Be_inversion_all} we have added also 
the $1/2^+$ LIT and response function used for the final evaluation of the
cross section, corresponding to the cutoff $\Lambda_3=400$ MeV.

\begin{figure*}[htbp!]
	\includegraphics[width=1.7\columnwidth]{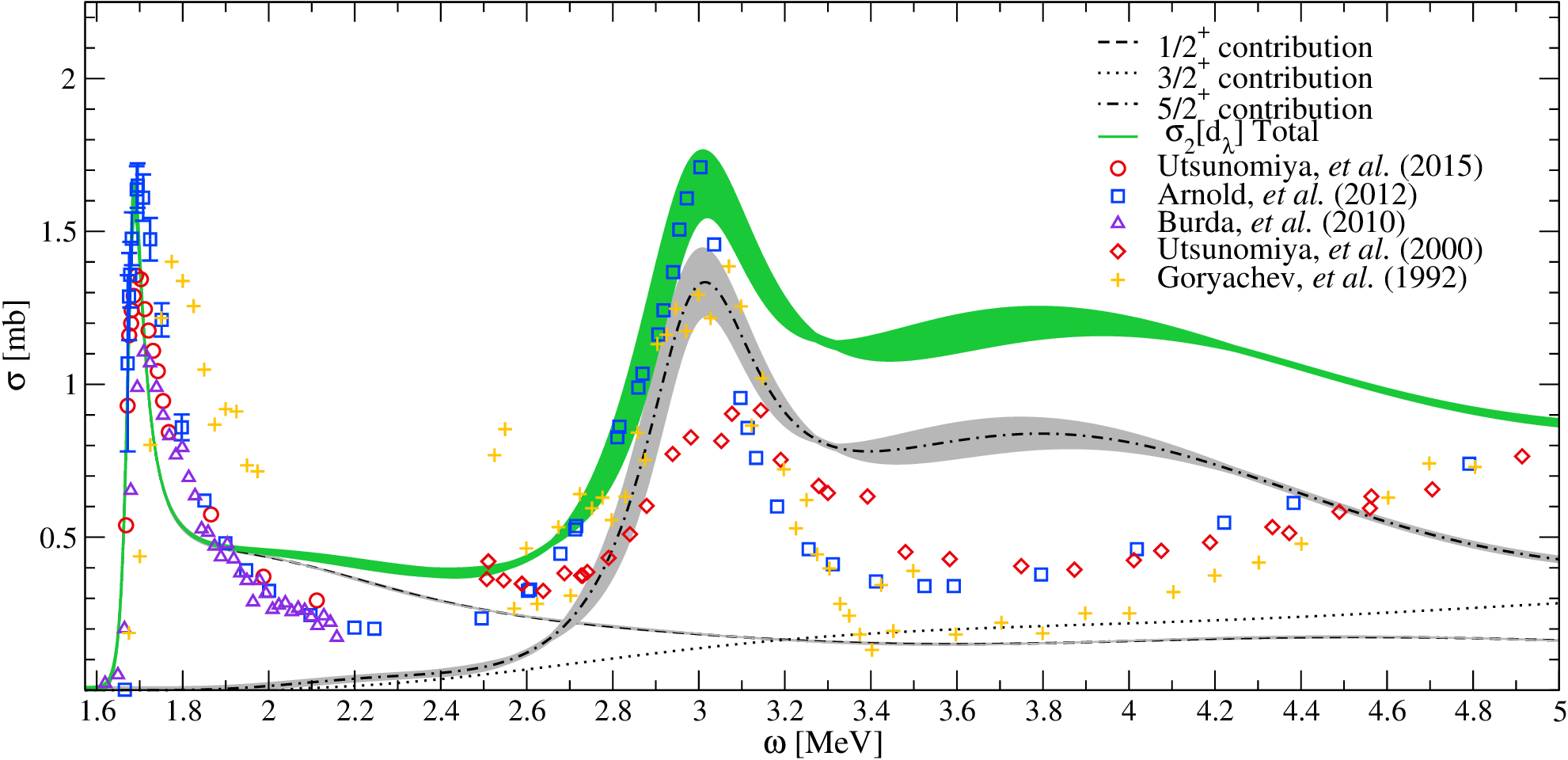}
	\caption{\label{fig:9Be_cs_total}%
		(Color online) Calculated low-energy $\Be$ photodisintegration cross
		section as a function of the photon energy $\omega$.
		$1/2^+$, $3/2^+$ and $5/2^+$ individual contributions are visible. 
		The grey band of the $1/2^+$ result represents the error due to the
		inversion procedure, while the $5/2^+$ band is related to the HH
		model-space convergence (see text for the details).
		The sum of the different contributions is depicted in green.  
		Experimental data from Refs.~\cite{utsunew,arnold,Burda,utsuold,Goryachev} 
		are also included.
	}
\end{figure*}

The $5/2^+$ and $3/2^+$ contributions to the $\Be$ photodisintegration cross
section are shown individually in Fig.~\ref{fig:9Be_cs_total},%
where the $1/2^+$ resonance of Fig.~\ref{fig:9Be_cs_12} 
with $\Lambda_3=400$ MeV has also been included to get the full picture.
For the $1/2^+$ and $5/2^+$ results, the associated errors are also shown. 
In the first case, as already mentioned, the error comes from the inversion
procedure; in the latter case, it is mainly due to the convergence in
$K_{\text{max}}$.
The total cross section can be obtained by summing the three individual
contributions. Different sets of experimental data
points available from literature are also shown to facilitate discussion of the
results. 
Near the three-body threshold at $1.573$ MeV and up to an energy of about 
$2$ MeV the entire cross section is given by the $1/2^+$ resonance. 
Then, the $5/2^+$ contribution starts to grow, peaking at $\approx 3$ MeV. 
As already mentioned, the resonance due to the $3/2^+$ state is very broad,
with a maximum contribution of $0.28$ mb at energies approaching $5$ MeV.
A clear overestimation of the experimental data in the range above $3$ MeV 
is evident from inspection of the plot. 
This is mainly due to the tail of the calculated $5/2^+$ peak.
An improvement of the calculations in the energy range beyond $3$ MeV is
probably obtained by including EFT terms beyond lowest orders. Particularly
important could be the shape parameter in the $\alpha n$ $P$-wave interaction.

%===========================================================  RESULTS-J1B
\subsection{\label{subsec:cross_jconv}Comparison of the dipole results with the one-body current calculations}
%========================================================================
In the following we introduce the results relative to the
$\Be$ photodisintegration cross section obtained with the one-body convection
current operator $\jmath_\lambda$ of Eq.~(\ref{eq:j}).

\begin{figure}[hbp!]
	\includegraphics[width=\columnwidth]{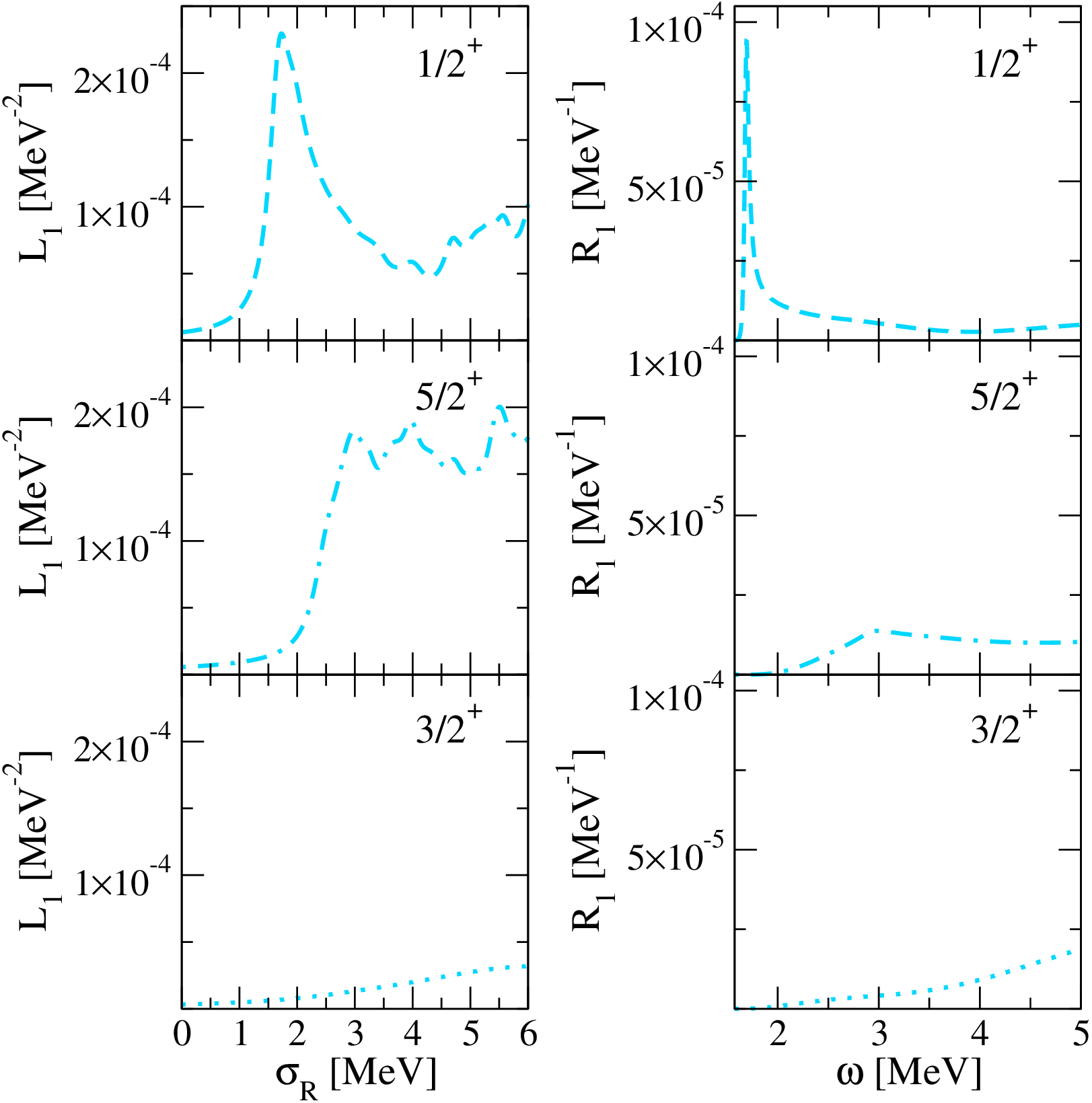}
	\caption{\label{fig:9Be_inversion_all_j}%
		Same as in Fig.~\ref{fig:9Be_inversion_all} but here the $1/2^+$,
		$5/2^+$ and $3/2^+$ LITs and response functions are calculated 
		by means of the convection current operator.
	}
\end{figure}

To allow a full comparison with the dipole cross section results previously
presented in Secs.~\ref{subsec:12+} and~\ref{subsec:total}, 
the same Hamiltonian is also employed in the one-body current calculations.
More specifically, the parameters of the two- and three-body potentials are
kept the same, as well as the parameters of the HH basis. 
Using Fig.~\ref{fig:9Be_inversion_all} as a reference, the corresponding
integral transforms $L_{1}(\sigma_R,\sigma_I)$ for the $1/2^+$, $5/2^+$ and
$3/2^+$ channels are represented in the left panels of
Fig.~\ref{fig:9Be_inversion_all_j},%
together with the respective response functions $R_{1}(\omega)$, 
which are visible on the right. The latter have been obtained by using the
procedure already explained in Sec.~\ref{subsec:12+}.
With each $R_{1}$, the individual $1/2^+$, $5/2^+$ and $3/2^+$ cross sections
can then be calculated by means of Eq.~(\ref{cross}).

\begin{figure}[hbp!]
	\includegraphics[width=\columnwidth]{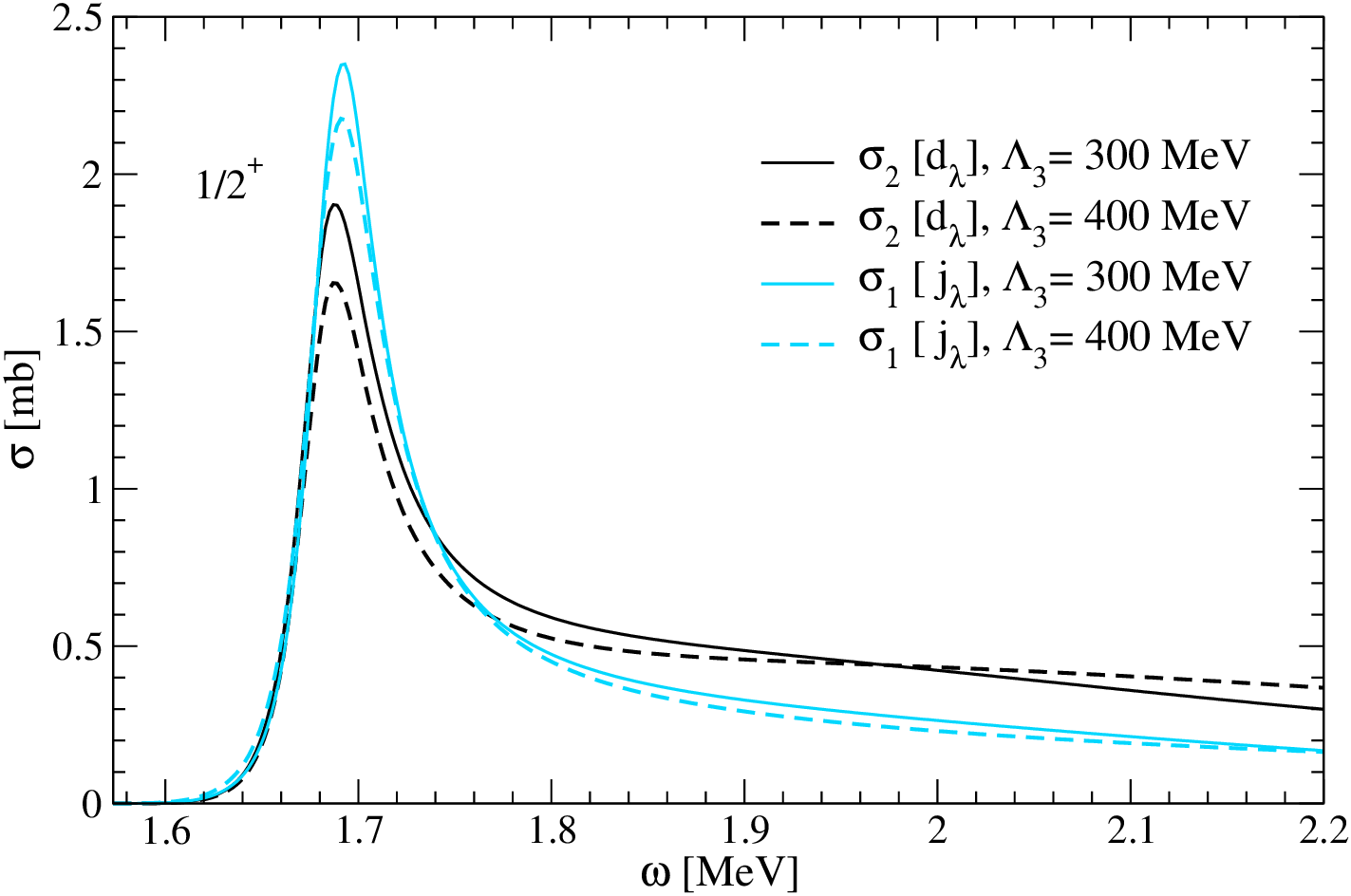}
	\caption{\label{fig:9Be_cs_12_d_j}%
		(Color online) Comparison between the $1/2^+$ contributions to the
		$\Be$ photodisintegration cross section calculated with the dipole
		operator $d_\lambda$ (black lines) and those obtained with the one-body
		current $\jmath_\lambda$ (light-blue lines). Results from computations
		with three-body cutoffs $\Lambda_3=300$ MeV (solid) and $\Lambda_3=400$
		MeV (dashed) are shown.
	}
\end{figure}

In Fig.~\ref{fig:9Be_cs_12_d_j}, the $1/2^+$ contribution to the 
$\Be$ photodisintegration cross section calculated with the dipole operator
$d_\lambda$, which has been already shown in Fig.~\ref{fig:9Be_cs_12}, 
is directly compared with the $\jmath_\lambda$-calculation.%
In the latter case, the peak position results to be slightly shifted by about 
$10$ keV. Moreover, the peak obtained with $\jmath_\lambda$ is $\approx 20\%$
and $30\%$ higher for fixed three-body cutoffs $\Lambda_3=300$ MeV and
$\Lambda_3=400$ MeV, respectively. However, the major difference between the
two calculations emerges above the energy of $1.9$ MeV. This should be entirely
ascribed to a non-vanishing contribution of the many-body currents, which is
not taken into account by the one-body current operator.

\begin{figure}[htbp!]
	\includegraphics[width=\columnwidth]{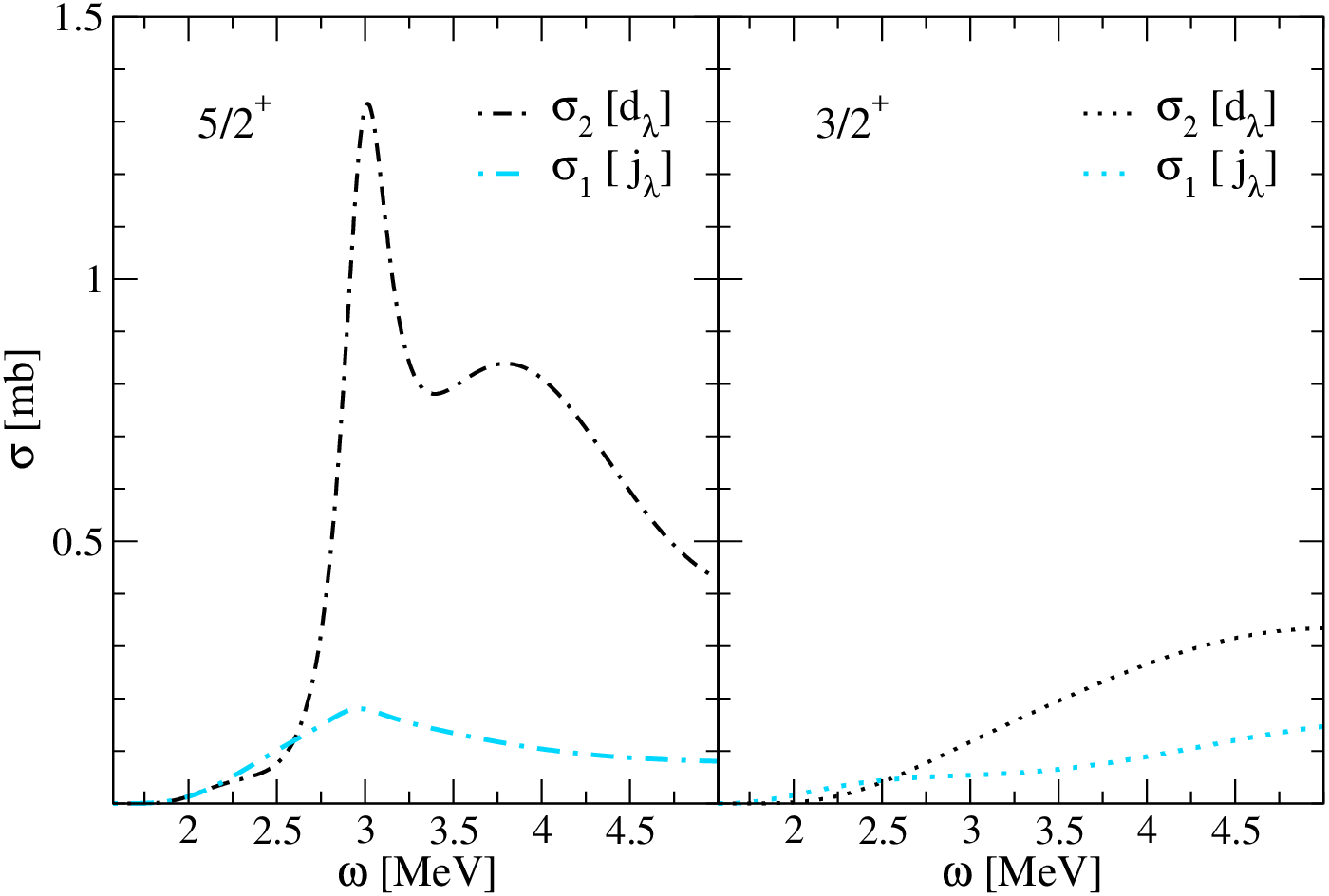}
	\caption{\label{fig:9Be_cs_52_32_d_j}%
		Same as in Fig.~\ref{fig:9Be_cs_12_d_j} but for the $5/2^+$ 
		(left panel) and $3/2^+$ (right panel) channels. 
		The three-body cutoff is fixed at $\Lambda_3=400$.
	}
\end{figure}

Finally, in Fig.~\ref{fig:9Be_cs_52_32_d_j},%
the $5/2^+$ and $3/2^+$ channels are also analyzed.
Interestingly, the two calculations of the $5/2^+$ contribution to the 
$\Be$ photodisintegration cross section give very different results: 
the resonance emerging around $3$ MeV from the $d_\lambda$-calculation, 
almost disappear when the one-body current is used. 
In fact, in the $\jmath_\lambda$-case the maximum value of the computed cross
section is $0.18$ mb.
A similar situation occurs in the case of the $3/2^+$ channel. 
The contribution to the cross section given by the one-body current calculation
is at most $0.15$ mb around $5$ MeV, 
less than half of the corresponding $d_\lambda$-calculation.
These results are a clear indication that the many-body currents 
play an important role in these channels.

\begin{figure*}[htbp!]
	\includegraphics[width=1.7\columnwidth]{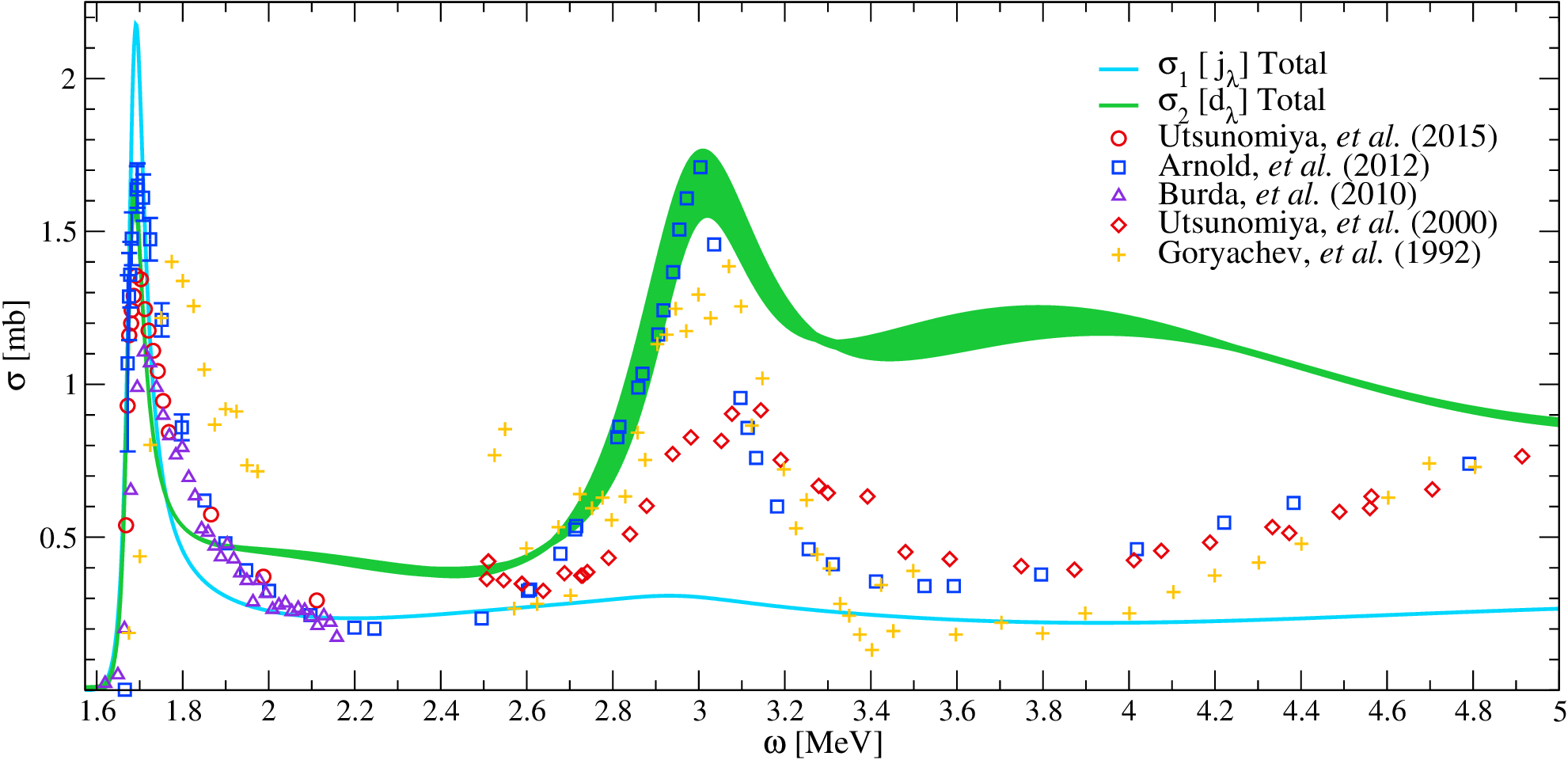}
	\caption{\label{fig:9Be_cs_total_d_j}%
		(Color online) Same as in Fig.~\ref{fig:9Be_cs_total} but here the
		difference between the total cross section calculated with the dipole
		operator $d_\lambda$ (green) and the one-body convection current
		$\jmath_\lambda$ (solid light-blue line) is highlighted.
	}
\end{figure*}

The sum of the individual $1/2^+$, $5/2^+$ and $3/2^+$ contributions to the
$\Be$ photodisintegration cross section as resulting from the one-body
convection current calculation is shown in Fig.~\ref{fig:9Be_cs_total_d_j}%
in light-blue. For ease of comparison, we have also included the
dipole-calculation (green) as well as the experimental data points.
Although the use of the operator $\jmath_{\lambda}$ leads to a $1/2^+$
resonance peak in the cross section that is very visible and slightly
overestimates both the $d_{\lambda}$-result and the experimental data, 
it seems to fail to reproduce the resonances occurring at higher energies.

%============================================================ Conclusions
\section{\label{sec:conc}Summary and conclusions}
%========================================================================
We have calculated the low-energy $\Be$ photoabsorption cross section in
an $\alpha\alpha n$ cluster approach considering exclusively electric dipole 
transitions. The electromagnetic transition operator has been considered
in two different ways: one-body convection current only, unretarded
dipole operator. The latter includes effects from many-body currents
(Siegert theorem).
A non-vanishing but very narrow contribution to the experimental low-energy
cross section is present around $2.4$ MeV~\cite{TILLEY2004155} 
in the $5/2^-$ channel.
It is due to a magnetic dipole transition, however, such a transition is not
included in our calculations, as it would be the subject of a different study.

The interactions among the cluster particles are derived in lowest order
Cluster EFT. Accordingly, we have taken into account the following partial
waves for the two-body interactions: $S$-waves for $\alpha\alpha$ 
and $\alpha n$, and the $P_{3/2}$ wave for $\alpha n$. 
For the potentials in any partial wave we have used a representation in
momentum space with a separable ansatz, 
which contains two low-energy constants and a Gaussian cutoff. 
The two LECs have been determined from an analytic solution of the
Lippmann-Schwinger equation. In fact such a solution can be compared to a
second-order effective range expansion containing scattering length and
effective range. We should mention that the $\alpha\alpha$ $S$-wave case is
more complicated because of the presence of the long-range Coulomb force. 
Here we have split the $T$-matrix into two pieces: a Coulomb part and a
Coulomb-distorted strong interaction part. 
In a final step, using experimental values for the scattering length and
effective range, we have determined the LECs. 
In addition to the two-body potentials, an $\alpha\alpha$$n$
hypermomentum-dependent three-body force has been used, 
again with a Gaussian cutoff.
	
For the three considered two-body interactions we have studied the cutoff
dependence and compared the calculated theoretical phase shifts with the
experimental ones. Similar checks have been made for the $^9$Be ground-state
energy. Concerning the phase shifts, rather satisfactory results have been
obtained in the energy range of interest, whereas the $^9$Be ground-state
energy depends rather strongly on the chosen cutoffs of the two resonant
partial waves ($\alpha\alpha$ $S_0$, $\alpha$$n$ $P_{3/2}$). 
The correct bound-state energy has then been obtained 
by  proper adjusting  the additional three-body potential.
	
The low-energy $^9$Be photodisintegration leads to an $\alpha\alpha n$ 
continuum state. Instead of calculating such $\alpha\alpha n$ states, 
we have used the LIT method, which reduces a continuum-state problem to a
bound-state-like problem. For the calculation of the LIT states, as well as for
the $^9$Be ground-state wave function, we have used expansions in
non-symmetrized hyperspherical harmonics in momentum space. 
In both cases the convergence of the expansion is satisfactory.

The $\alpha n$ $S$-wave interaction plays an interesting role. 
It has a negligible effect on the $^9$Be ground-state energy, but affects
quite  significantly the cross section of the 1/2$^+$ resonance, 
located closely  beyond the breakup threshold. 
For energies above the 1/2$^+$ resonance its role is again  marginal.
	
It has been necessary to take a state-dependent three-body potential. 
In fact a state-independent one leads to the wrong positions of the 1/2$^+$ 
and 5/2$^+$ resonances. Therefore, the strength of the three-body force has
been adjusted to get the correct resonance positions. 
Nonetheless, resonance peak heights and widths in the photoabsorption cross
section are predictions of the present theory. The comparison with experimental
data has shown a good agreement in the peak heights for both resonances. The
situation regarding the width is somewhat different: the width of the 5/2$^+$
resonance is quite consistent with the distribution of the experimental data,
while the 1/2$^+$ resonance width is somewhat too small. Additionally, for the
5/2$^+$ case, the cross section beyond the resonance is much too large.
	
In the $1/2^+$ calculation, the one-body convection current operator 
leads to slightly different results with respect to the dipole.
An interesting effect is the large contribution of many-body currents for
the 5/2$^+$ resonance. In principle, this does not come as a  surprise, 
since the interaction in the $P_{3/2}$ partial wave has a strong momentum
dependence and enters at the lowest order of the EFT expansion.
The use of Siegert theorem ensures Gauge invariance, and this is broken already
at LO with pure one-body convection current calculations.

An improvement in the calculations at higher energies is likely to be achieved
by including higher-order $\a n$ and $\a\a$ terms beyond the lowest EFT orders. 
Of particular interest is the inclusion of the shape parameter in the $\a n$
$P_{3/2}$ potential. This should lead to a significant modification of the
effects from the many-body currents and may help to reduce the relatively large
cross section observed beyond the 5/2$^+$ resonance.
The contribution of the $\a n$ $P_{1/2}$ potential to the cross section could
also be explored. Anyway, this has been proven to be negligible for the $1/2^+$
resonance~\cite{Capitani}, i.e.~in the energy range just above the breakup
threshold.

\begin{acknowledgments}
Y.C. and W.L. acknowledge the financial support of 
the European Union - Next Generation EU, Mission 4 Component 1, 
CUP I53D23001060006, for the PRIN 2022 project 
``Exploiting separation of scales in nuclear structure and dynamics''.
C.J. was supported by the National Natural Science Foundation of China 
(Grant Nos. 12175083, 12335002, and 11805078).
Two of us (W.L. and G.O.) thank D. Phillips for helpful discussions.
This research was supported in part by grant NSF PHY-2309135 
to the Kavli Institute for Theoretical Physics (KITP).
The University of Trento computing resources are also acknowledged.
\end{acknowledgments}

\bibliography{9Be_biblio.bib}

\end{document}